\documentclass[conference]{IEEEtran}

\makeatletter
\def\ps@headings{%
\def\@oddhead{\mbox{}\scriptsize\rightmark \hfil \thepage}%
\def\@evenhead{\scriptsize\thepage \hfil \leftmark\mbox{}}%
\def\@oddfoot{}%
\def\@evenfoot{}}
\makeatother
\usepackage{graphicx}
\usepackage{subfig}
\usepackage{xspace}
\usepackage{epsfig}
\usepackage{amsmath}
\usepackage{multirow}
\newcommand{\etal}{{\em et al.}}
\newcommand{\ignore}[1]{}
\newcommand{\figref}[1]{Fig.~\ref{#1}}
\newcommand{\tabref}[1]{Table~\ref{#1}}

\newcommand{\figsize}{0.45}

\begin{document}
\title{Topological Trends of Internet Content Providers}

\author{
\IEEEauthorblockN{Yuval Shavitt}
\IEEEauthorblockA{School of Electrical Engineering\\
Tel-Aviv University, Israel\\
Email: shavitt@eng.tau.ac.il}
\and
\IEEEauthorblockN{Udi Weinsberg}
\IEEEauthorblockA{\IEEEauthorblockA{School of Electrical Engineering\\
Tel-Aviv University, Israel\\
Email: udiw@eng.tau.ac.il}}
}

\maketitle

\begin{abstract} The Internet is constantly changing, and its hierarchy was recently shown to become flatter. 
Recent studies of inter-domain traffic showed that
large content providers drive this change
by bypassing tier-1 networks and reaching closer to their users, enabling them to save transit costs
and reduce reliance of transit networks as new services are being deployed, and traffic shaping is becoming
increasingly popular. 

In this paper we take a first look at the evolving connectivity of large content provider
networks, from a topological point of view of the autonomous systems (AS) graph.
We perform a 5-year longitudinal study of the topological trends of large content providers, by analyzing several
large content providers and comparing these trends to those observed
for large tier-1 networks. We study trends in the connectivity of the networks, 
neighbor diversity and geographical spread, their hierarchy, the adoption of IXPs as a convenient method for peering,
and their centrality. Our observations indicate that content providers gradually increase
and diversify their connectivity, enabling them to improve their centrality in the graph, and as a result,
tier-1 networks lose dominance over time. 
%Using selective peering agreements,
%using their own networks for transit and increased usage of exchange points, large content providers increase their centrality.
\end{abstract}

\section{Introduction}
\label{introduction}
The Internet is a constantly evolving network, quickly adapting to customer needs and
financial forces.
Up until recently
it was common to picture an hierarchical Internet~\cite{ge01hierarchical,dimitropoulos-2006-37,medusa-pnas}, in which networks are
either tier-1, large networks that provide global transit functionalities,
tier-2, smaller
Internet Service Providers (ISPs) that provide Internet connectivity to
their customers, or stub networks that produce and consume
content~\cite{tiers}.

However, in recent years the Internet is changing. The appearance and rapid growth
of large content providers, such as Google, Yahoo! and others, is gradually changing the 
roles of key Internet players to accommodate their needs. First, large content providers produce huge
amount of content that is consumed by users around the globe, inducing heavy traffic on tranist networks.
Although wholesale transit prices are decreasing by roughly 30\% each year~\cite{peer}, and transit
providers offer various wholesale pricing plans to accommodate these needs, such as tiered prices~\cite{tiers},
content providers still seek ways to significantly cut transit costs. Furthermore, as content providers 
deploy an increasing number of Software as a Service (SaaS), such as elastic computing, collaboration tools, storage
and even complete content delivery networks (CDNs), they seek to reduce reliance on transit providers
that were reported to perform traffic shaping~\cite{beverly2007,amogh2008}.

\ignore{
Previous papers made various assumptions regarding the way these content
provider networks are operated. For example, attempting to reduce costs
was assumed \cite{incomplete} to be achieved by peering with transit networks, large and small, identifying hot-spots
and quickly reacting to changes in the observed traffic. However most of these assumptions are
mostly learned guesses, and are not backed by real observations.

Moreover, although the rapid growth of these networks clearly effects other large players
in the Internet's ecosystem, this effect was never thoroughly studied.
}

As a result of these trends, the Internet was reported to be forming a flatter and denser
network~\cite{flatnet,labovitz,amogh2010}, mostly using observations of traffic flows. 
In this paper we take a first look at the changing connectivity of large content provider
networks, from a topological point of view. Unlike previous work
that studied various traffic characteristics \cite{sprint,maier2009,labovitz}, we consider
the trends observed in the connectivity of the networks in the Autonomous Systems (AS) graph.

We achieve this
using a 5-year longitudinal study of the AS graph, focusing on 5 major content providers: Google, Yahoo!, MSN,
Amazon, and Facebook. The first three are well established, large content providers, that
have been around before the beginning of our study, in 2006.
Amazon provides a unique opportunity to study a content provider that changed its
scope (from an online store to a cloud service host) and Facebook reveals
the high-paced growth of an extremely popular content
provider. Using a comparative approach, we examine 5 major transit providers, namely AT\&T, Qwest, Level3,
Sprint, and Global Crossing (Glbx). All of these are large, tier-1 transit networks \cite{ground-truth}, that have been used
by content providers for transit services over the years.

In this paper we create a snapshot of the AS-level graph every 3 months, using a month of active traceroutes, from late 2006 till early 2011. 
We then
study the connectivity trends, meaning, how content and transit providers are connected and evolve over time. We look
at the number of neighboring ASes, types of networks they connect with and spatial spread for understanding these trends. We
then look at the adoption of IXPs which are a convenient and cost-effective method for peering between ASes. We then study
the changes in the hierarchal position of content providers and conclude with studying their centrality.

Understanding the evolving trends of AS connectivity has implications on different aspects of the Internet ecosystem.
The decreasing dominance of large transit providers we observe indicates a change in the way traffic flows and
networks interconnect. These in turn have direct implications on
the operational decisions that drive ASes, their connectivity and profitability. Additionally, understanding these trends
can help improve Internet research, such as growth models \cite{amogh2010} and traffic flow analysis \cite{labovitz}.

\section{Related Work}
Several recent papers study the emerging change in the Internet ecosystem, which is driven mostly by large scale content providers.
One of the early observations of this change was made by Gill~\etal~\cite{flatnet}. The authors showed that large content providers
bypass many tier-1 ISPs by pushing their networks closer to the users, and suggested that such a trend can possibly flatten
the Internet.

Kuai \etal~\cite{596565}, He \etal~\cite{HeSFK07,lordoflinks}, and more recently Augustin~\etal~\cite{ixpsmapped} studied the AS-graph,  and
discuss in details methods for
discovering IXP participants. These works
report significantly higher number of peering relationships
discovered among ASes that are IXP participants than among ASes that
are not connecting via an IXP.

Dhamdhere and Dovrollis \cite{amogh2010} presented a new Internet model that captures the
Internet transition from a hierarchy of transit providers to a flatter interconnection of peers.

Most recently, Labovitz~\etal~\cite{labovitz} performed a large-scale two-year study of the inter-domain
traffic, showing that the amount of traffic originated from content providers is rising, and most of it
is routed outside of the traditional Internet core. Specifically, more than 5\% of all inter-domain
traffic in July 2009 originated from Google's networks. Additionally, they showed that content providers
often use their own networks for saving transit costs. As an example, YouTube inter-domain traffic was shown
to decay rapidly by merging it into Google's own peering and data centers.
%This observation strongly validates an assumption made by previous work \cite{incomplete}.

\begin{figure*}[tbh]
\centering
    \subfloat[Content]{
	\label{fig:degree_content}
    \epsfig{file=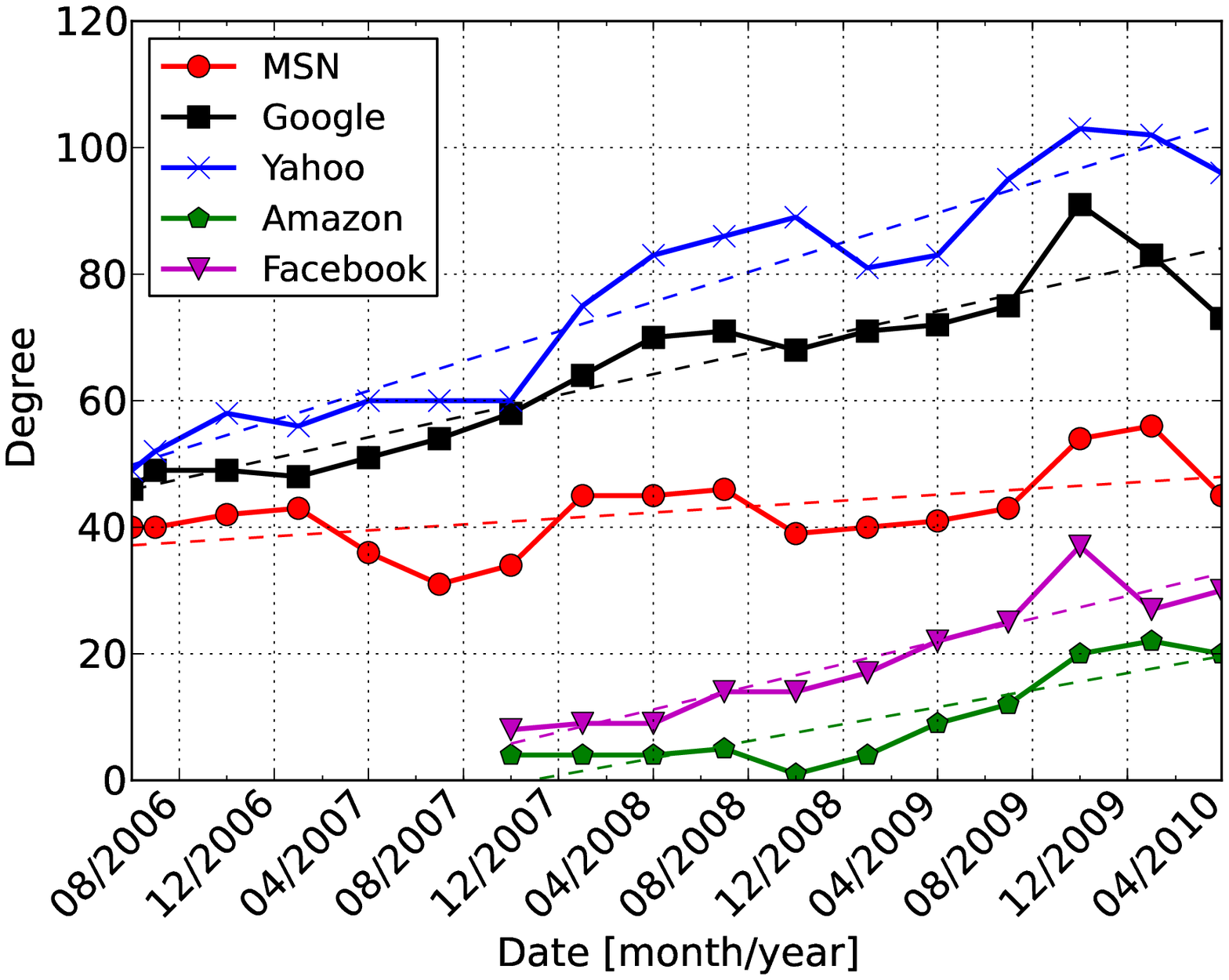,width=\figsize\textwidth}
    }
    %\hspace{-1mm}
    \subfloat[Content neighbors]{
	\label{fig:degree_content_neighbors}
    \epsfig{file=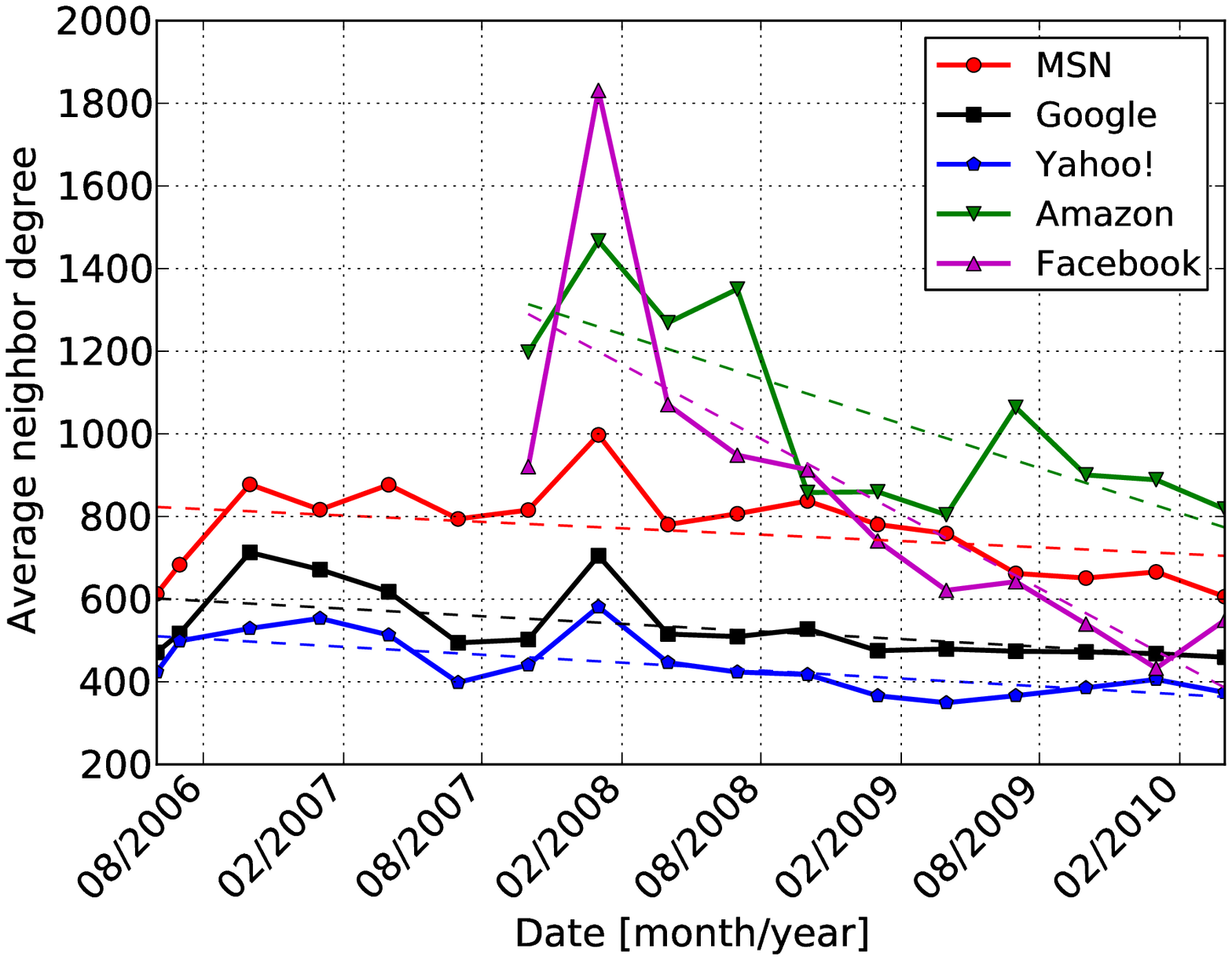,width=\figsize\textwidth}
    }
    \hspace{-1mm}
    \subfloat[Transit]{
    \label{fig:degree_transit}
    \epsfig{file=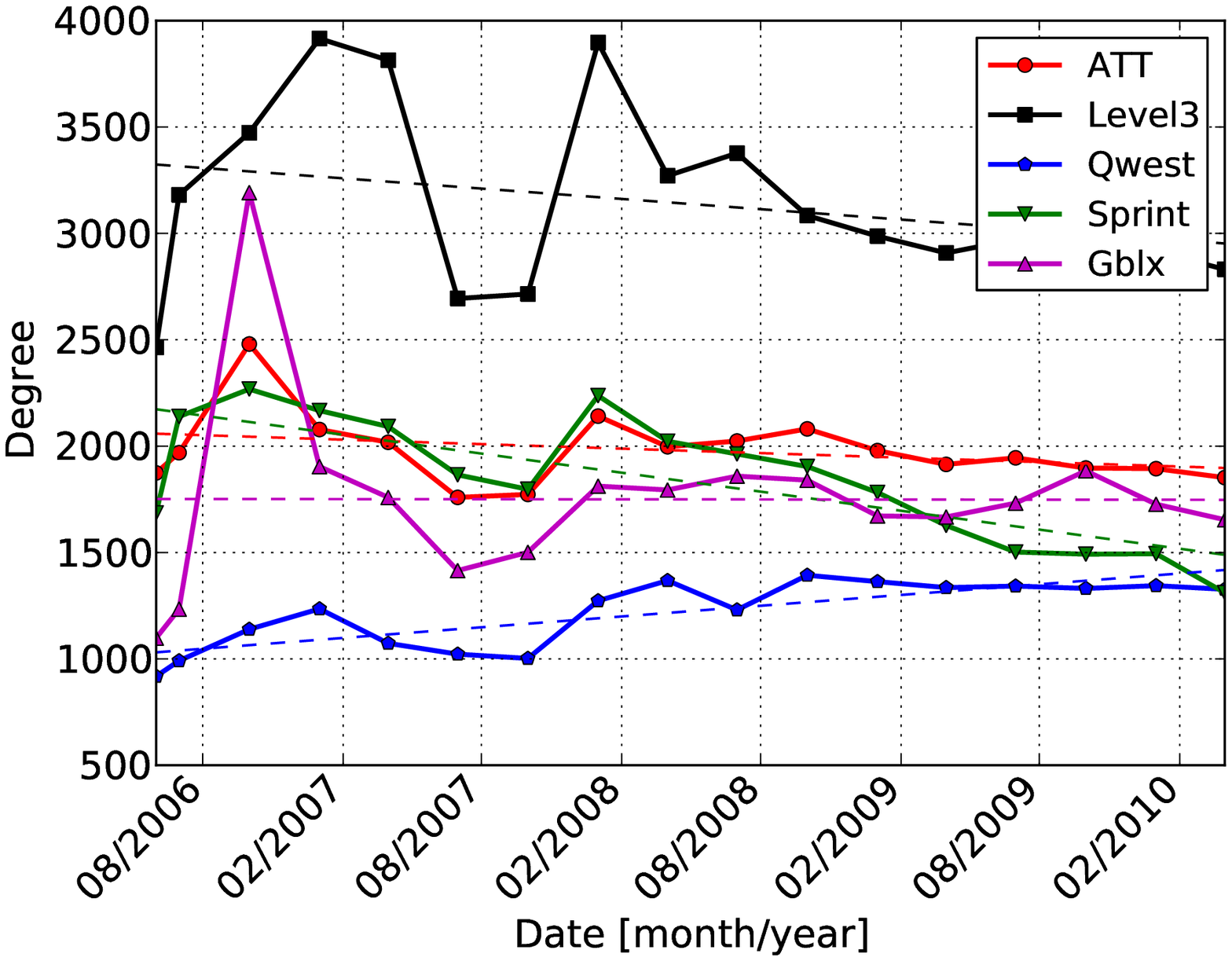,width=\figsize\textwidth}
    }
    %\hspace{-1mm}
    \subfloat[Transit neighbors]{
    \label{fig:degree_transit_neighbors}
    \epsfig{file=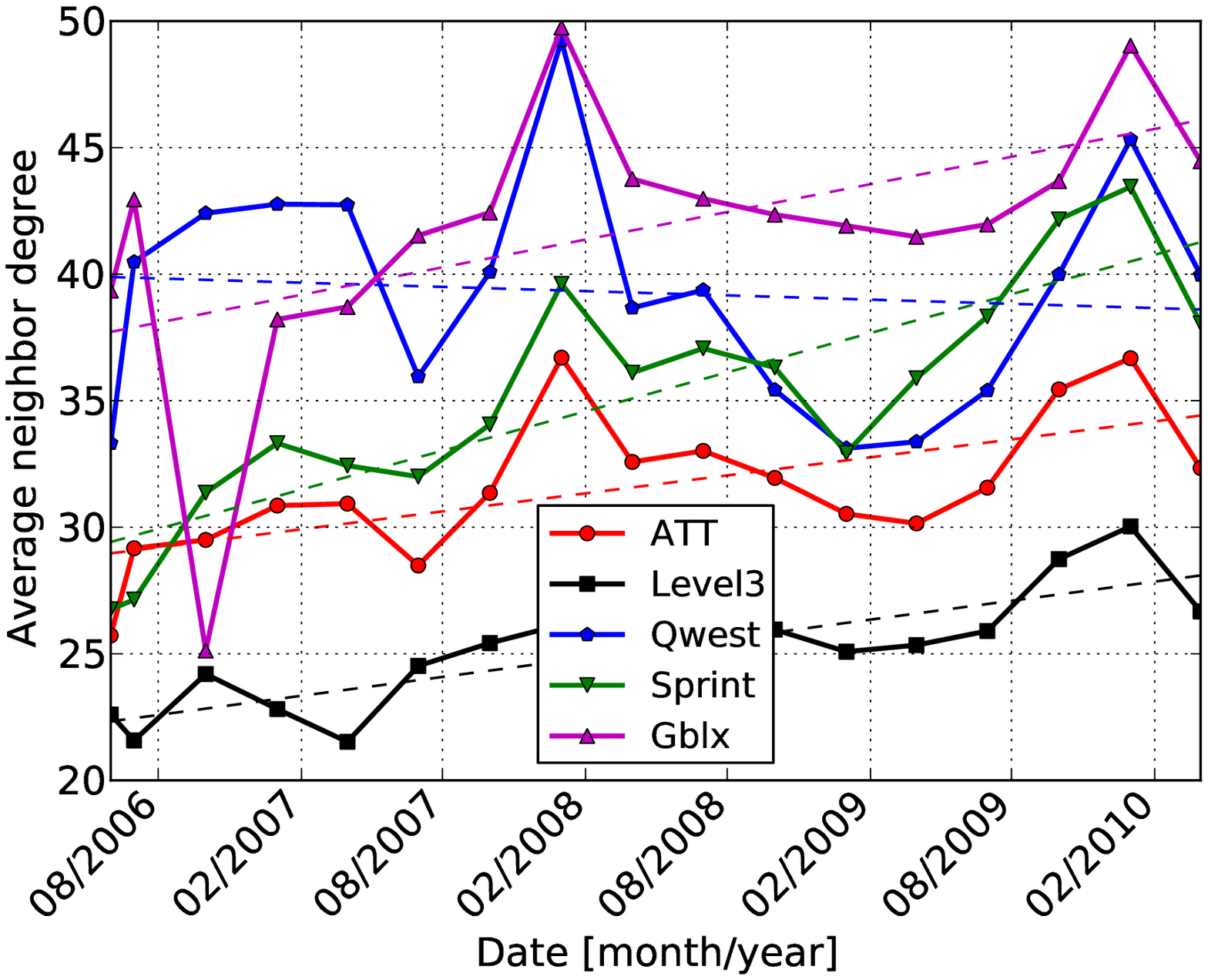,width=\figsize\textwidth}
    }
    \caption{The degree of transit and content networks, and their average neighbor degree in the AS-graph}
    \label{fig:degree}
\end{figure*}

\begin{figure*}[tbh]
\centering
    \subfloat[Neighbors types]{
	\label{fig:type_content}
    \epsfig{file=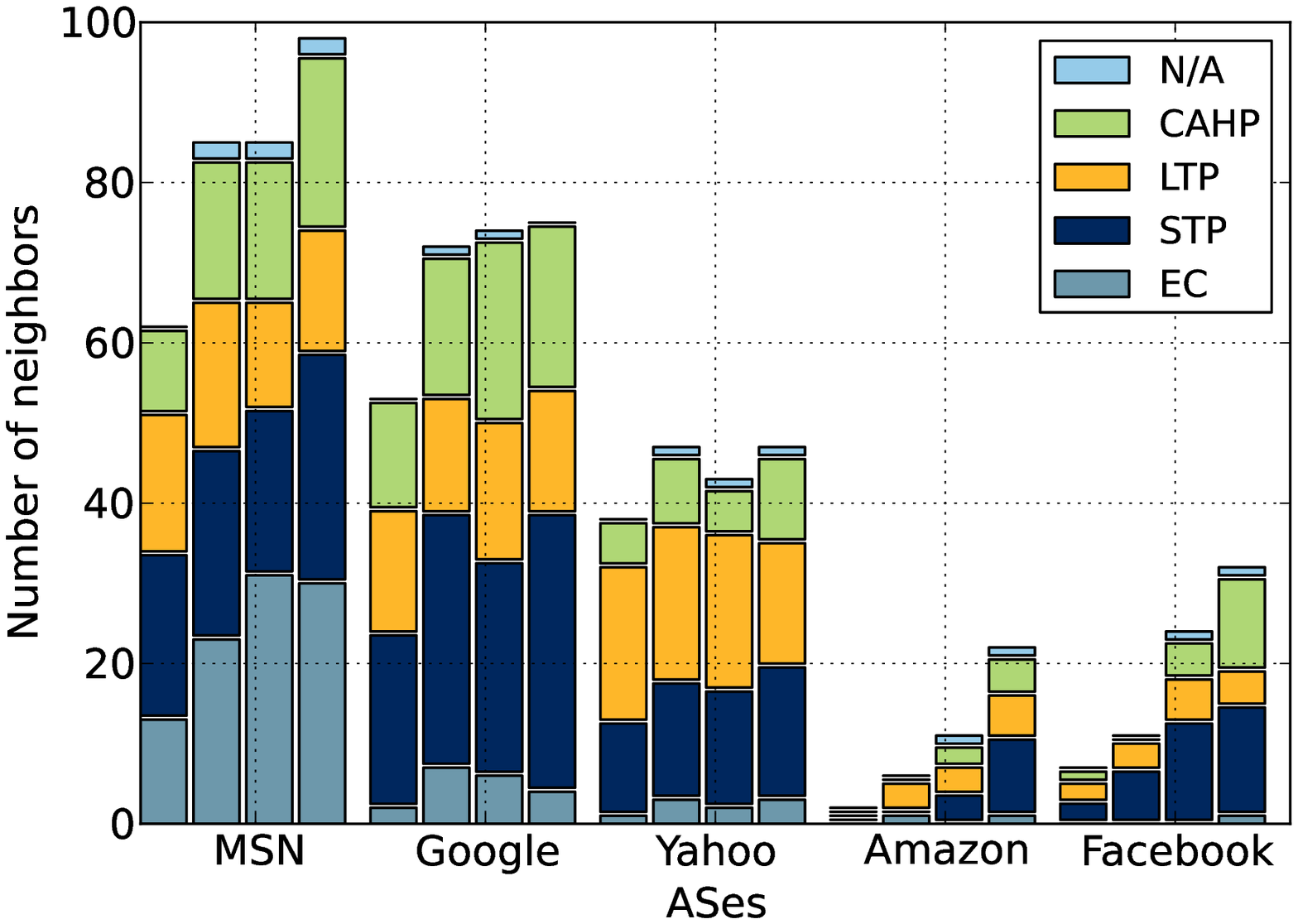,width=\figsize\textwidth}
    }
    \hspace{-1mm}
    \subfloat[Neighbors countries]{
	\label{fig:country_content}
    \epsfig{file=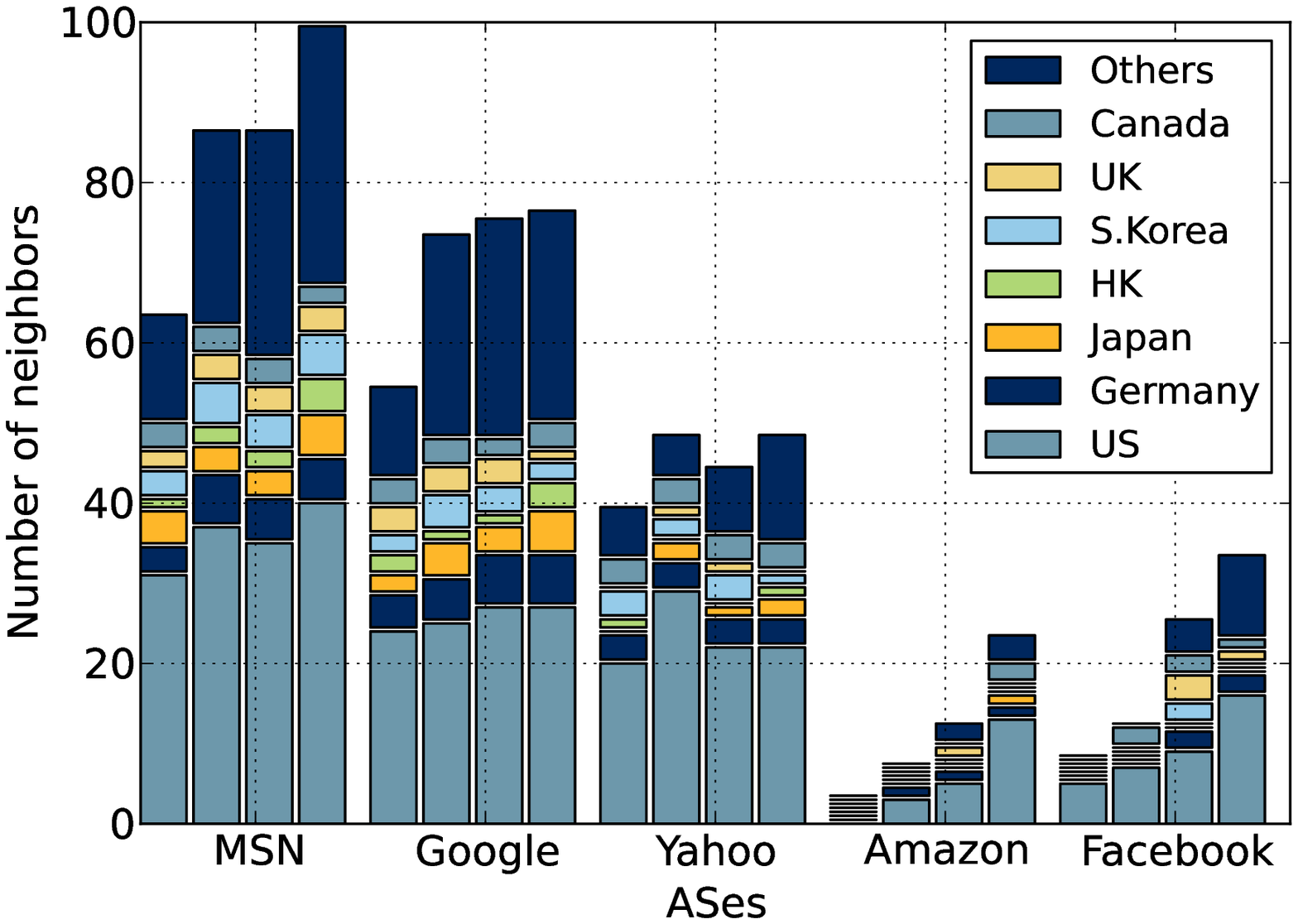,width=\figsize\textwidth}
    }
    \caption{Content providers neighbor types and countries}
    \label{fig:neighbor_types}
\end{figure*}

\section{Methodology}
For the purpose of this study we build the Internet AS graph 
every three months since January 2006 until July 2010. The AS graph is built by traversing
IP-level traceroutes and resolving each IP into its corresponding AS. We resolve IP to
AS using the published iPlane nano \cite{nano} database, which is a set of mappings between
IP prefixes into ASes, collected from all major BGP monitors. This set contains 326,102 prefixes
mapped to 30,779 ASes.
IPs that were mapped
into AS sets or multi-origin ASes (MOAS) were treated as unresolved.

%% TBD -- need to validate this in two ways:
% - IXP: Mapped? what do they say
% - Download a new IXP prefix table from PCH and see if things changed in a "bad" way - prefixes are removed?
% - Send an email to PCH operator, and see if they know of prefixes that were ASes and then converted to IXPs
Additionally,
we used Internet Exchange (IXP) prefix mapping provided from \cite{ixpsmapped}. The IXP list
is comprised of prefix lists collected from Packet Clearing House (PCH) \cite{pch}, PeeringDB \cite{peeringdb}
and additional manually collected sources.
This list provides us with 393 prefixes belonging to 278 IXPs worldwide. We 
use the same IXP list, obtained in late 2009 for all years, assuming that the assignment of prefixes to
IXPs did not decreased overtime, i.e., even if an IXP becomes defunct \cite{pch}, there are only a few or no cases
that a prefix, which was assigned to an IXP in 2009,
was assigned to a real AS in other times.

Each AS graph is built by traversing AS traces of a single month, creating a link between
two ASes that follow each other in the AS trace, or have an IP that belongs to an IXP prefix between them (the latter follows the technique described in \cite{lordoflinks} and extended in \cite{ixpsmapped}).

For the IP-level traces we use two datasets, DIMES \cite{dimes-ccr} and iPlane \cite{iPlane}.
DIMES is a community-based Internet mapping effort, measurement from thousands of vantage points, located
at user homes and since 2010 also in PlanetLab \cite{PlanetLab}. iPlane uses PlanetLab nodes and traceroute
servers, measuring from a relatively stable set of 300 servers. Although DIMES, due to its
diverse distribution, uncovers more links than iPlane \cite{quantify-jsac}, it suffers
from vantage point churn, making its observed topology more ``noisy" and susceptible to measurement
artifacts. iPlane on the other hand,
is more stable, both in the number of measuring vantage points and the target IPs, however its
topology is smaller.

In this paper we are interested
in global {\em trends} observed in the ecosystem of large Internet players. Since these networks are well
observed by both DIMES and iPlane, we expect both platforms to capture similar trends that
effect the Internet ecosystem, even if the exact numbers are somewhat different.

Therefore,
when analyzing trends of the entire AS-level graph, we use DIMES data, which
brings a more accurate view of the topology, while the noise gets smoothed due
to the large amount of data. When analyzing specific ASes, we
use iPlane's data, since it is more stable. Indeed, in most cases, both infrastructures resulted
in the same overall trends, assuring that the observations we make are indeed due to real topology and routing characteristics
and are not the results of some measurement bias. Whenever the two datasets do not agree, we
present both results and discuss the differences and their causes.

It is important to note that since we are interested in trends, the exact numbers we obtain (for number of connections, 
clustering coefficient, etc.) are not important.
We are interested in the their scale and especially in their evolution over time.  
Thus, the effect of measurement errors \cite{paris}, errors in the IP to AS translation \cite{mao03towards,mao04scalable,1064257}, and similar inaccuracies,   % add citations 
which are discussed in the measurement literature, are not significant for this study since they can not effect trends.

%In addition to the AS graph, we extract the observed usage of the networks in IXPs, in order
%to better understand their usage.

\section{Autonomous Systems Connectivity}
%In this section we study the connectivity of large content and transit providers in the AS graph.

The simplest way to measure the connectivity of an AS is the number of neighboring ASes it is connected to, i.e., its degree
in the AS graph. \figref{fig:degree_content} and
\figref{fig:degree_transit}
show the expected vast difference between the degree of content providers and transit networks.
However, while \figref{fig:degree_content} shows that content providers are increasing their connectivity over
time, the connectivity of transit networks depicted in \figref{fig:degree_transit} exhibits a slow decrease (except Qwest).

\figref{fig:degree_content_neighbors} shows that the average degree of the neighboring ASes of the
three `veteran' content
providers slowly decreases over time, while the new ones, Amazon and Facebook, start with a high average neighbor degree, which is quickly
reduced to match the other three. This indicates that the content providers start by connecting mostly to tier-1
providers, but as they expend they connect to additional providers which are mostly not tier-1 (see additional discussion below).
\figref{fig:degree_transit_neighbors} shows that the average neighbor degree of the transit networks increase. This
can be an indication that small customers are disconnecting from the transit networks, or alternatively, that the tier-1
neighbors are becoming better connected, hence increase their degree. Both reasons imply that transit networks
are loosing some of their dominance in Internet connectivity.

\ignore{
\begin{figure*}[tbh]
\centering
    \subfloat[Content]{
	\label{fig:degree_content}
    \epsfig{file=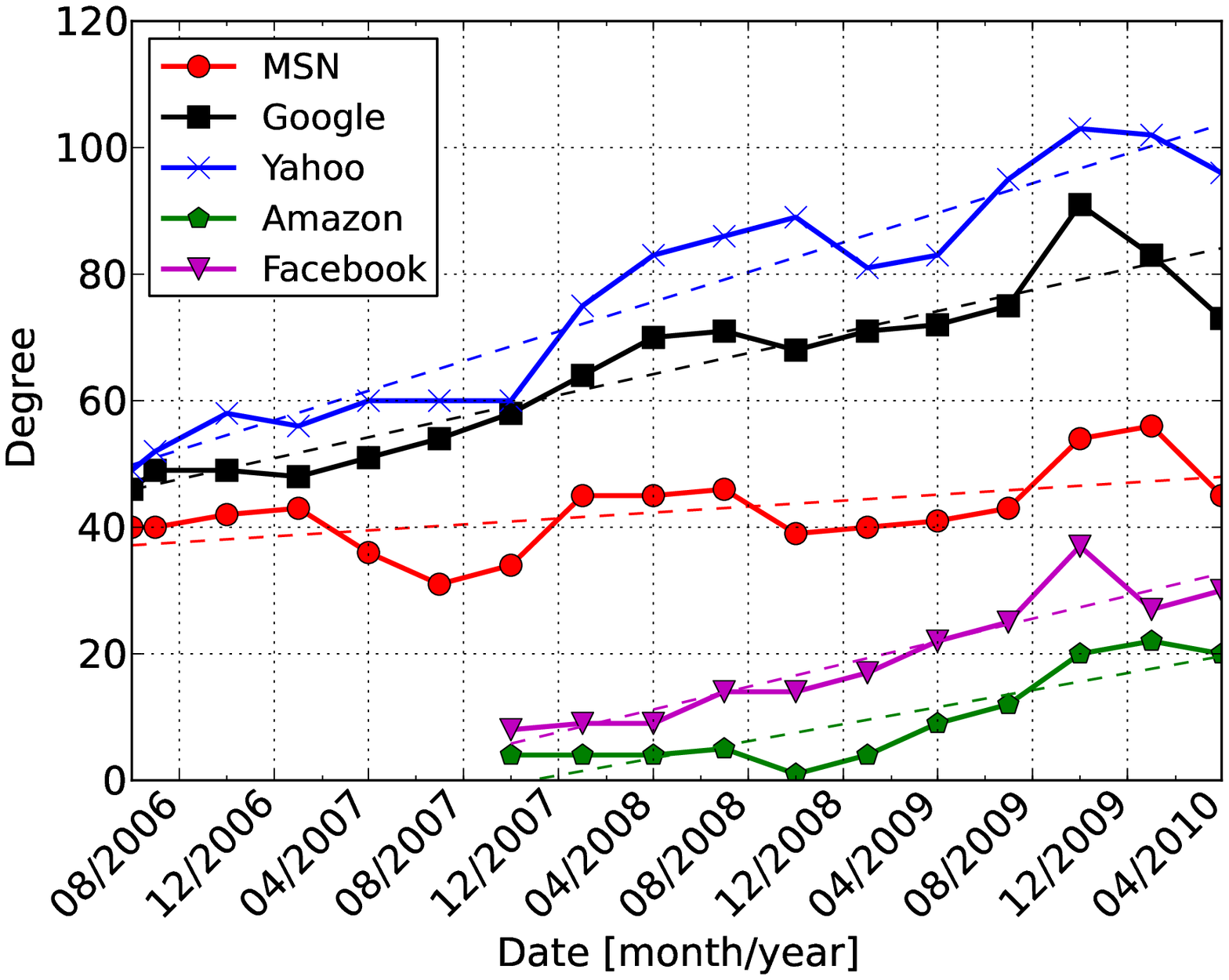,width=\figsize\textwidth}
    }
    %\hspace{-1mm}
    \subfloat[Content neighbors]{
	\label{fig:degree_content_neighbors}
    \epsfig{file=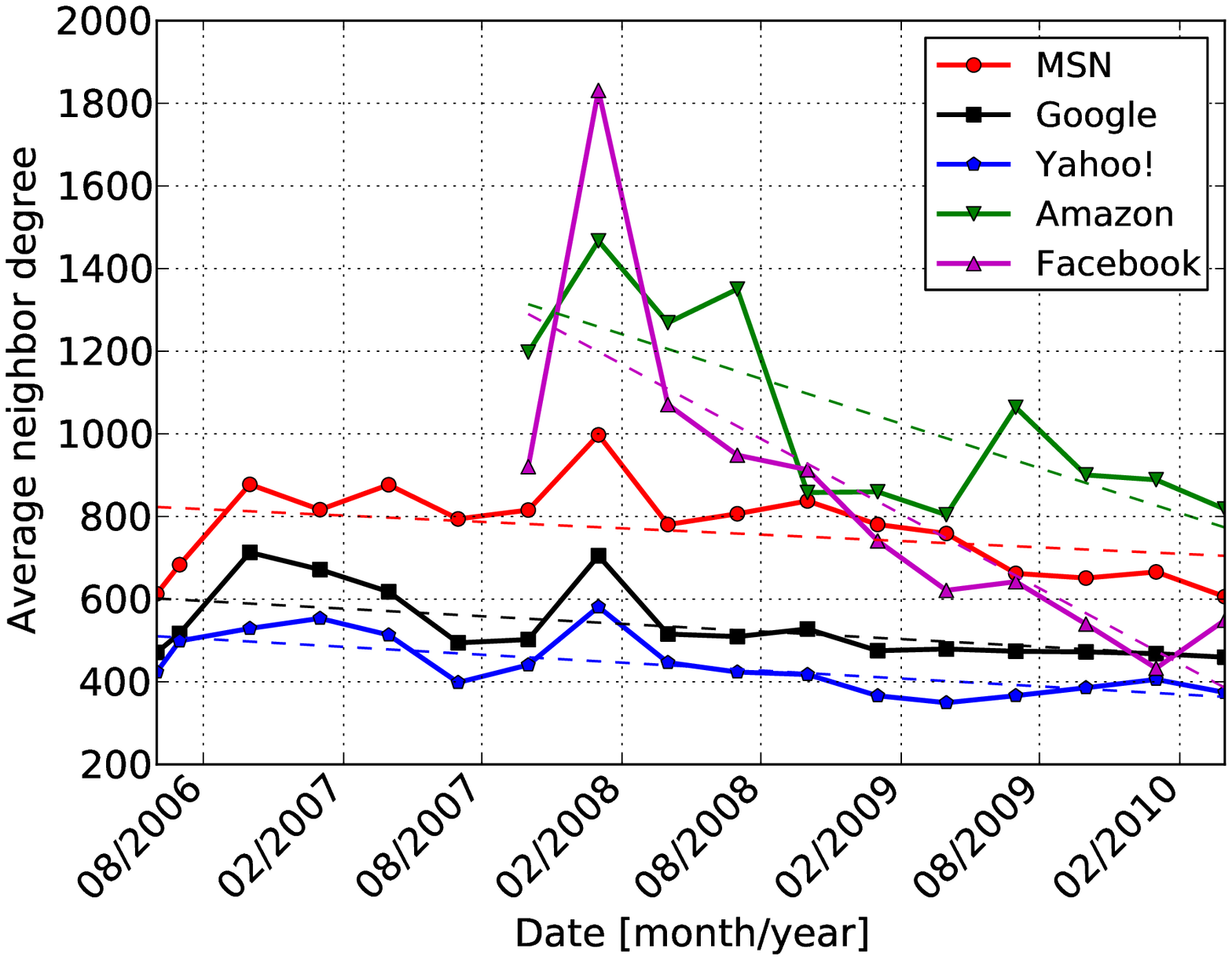,width=\figsize\textwidth}
    }
    \hspace{-1mm}
    \subfloat[Transit]{
    \label{fig:degree_transit}
    \epsfig{file=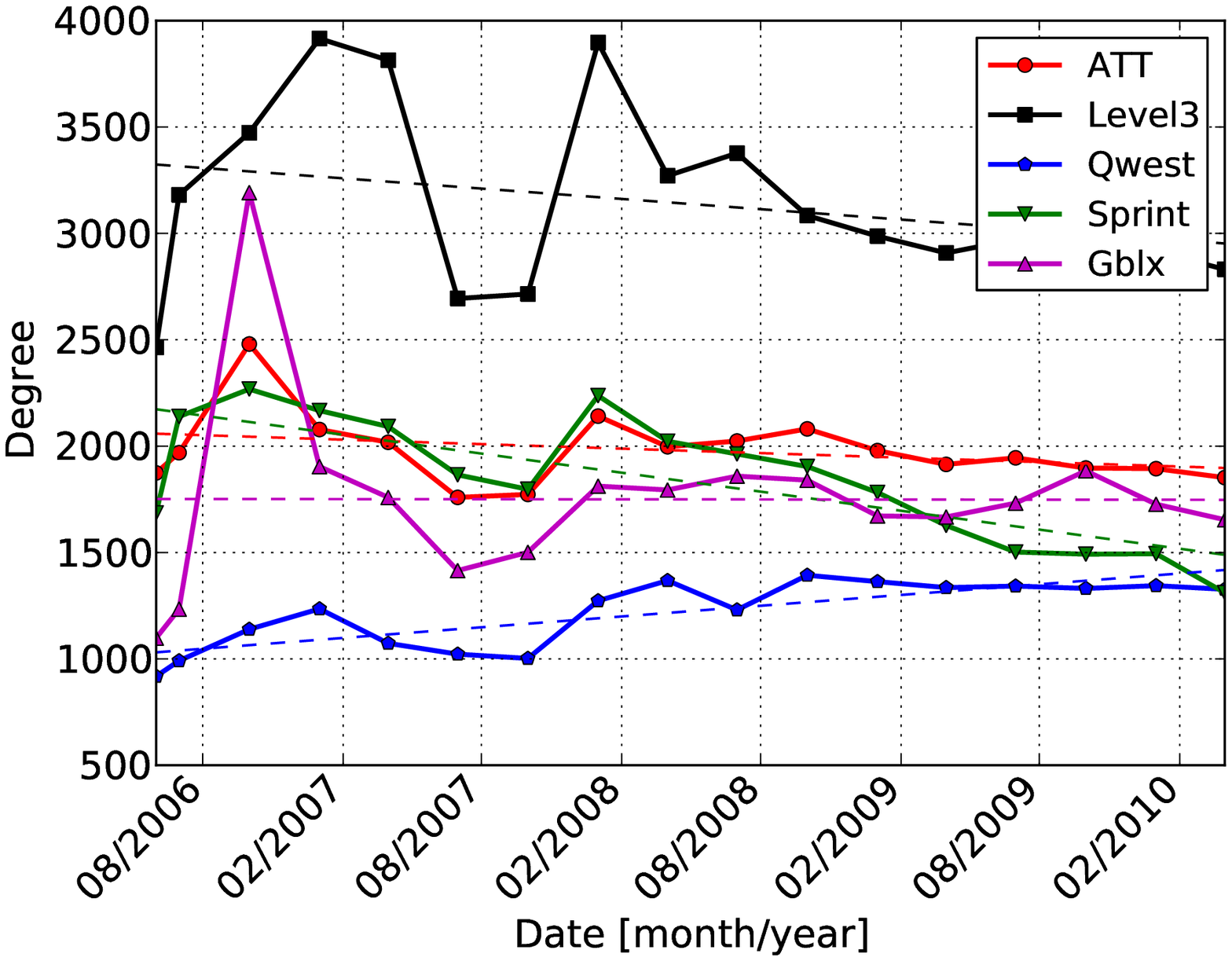,width=\figsize\textwidth}
    }
    %\hspace{-1mm}
    \subfloat[Transit neighbors]{
    \label{fig:degree_transit_neighbors}
    \epsfig{file=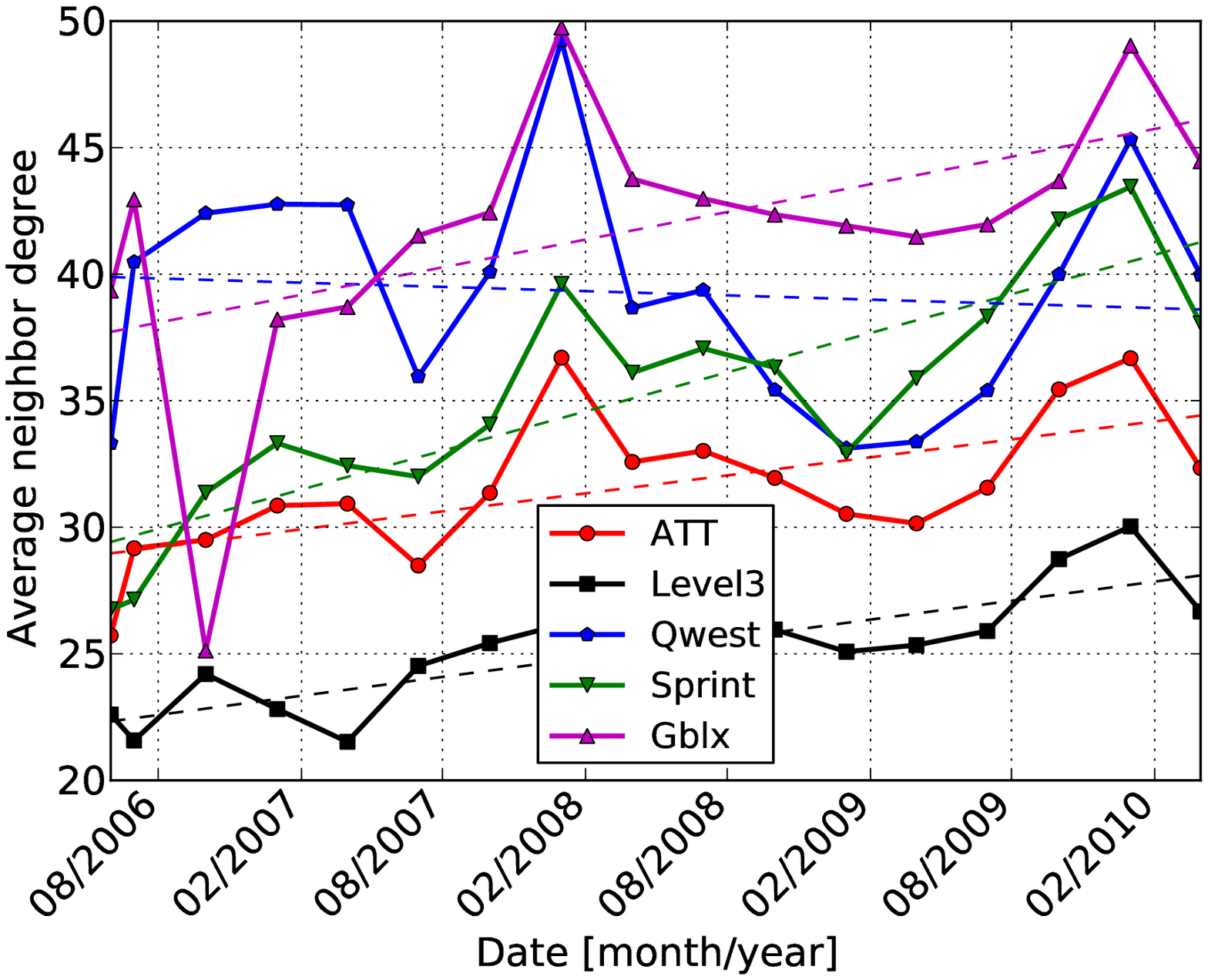,width=\figsize\textwidth}
    }
    \caption{The degree of transit and content networks, and their average neighbor degree in the AS-graph}
    \label{fig:degree}
\end{figure*}
}

To further understand the reasons behind these trends, we classify the neighbors of each AS.
We use the classification provided by Dhamdhere and Dovrolis \cite{tenyears}, in which
an AS is classified as an enterprize customer (EC), a small or large transit provider (STP and LTP), or a
content, access and hosting provider (CAHP). The authors base their classification on the average customer
and peer degrees of the AS over its entire lifetime within a 10 year longitudinal study,
and claim to reach over 80\% accuracy.

\figref{fig:type_content} depicts the distribution of the AS types of the neighbors of the content providers
as captured during April in four consecutive years: 2007 (left-most bar in each AS) till
2010 (right-most bars). The figure shows that the number
of LTP neighbors did not significantly change, however, since the number of neighbors increased over time for all content providers, the
overall LTP  percentage decreased. The most drastic
change is observed for Amazon and Facebook, with a $2/3$ reduction. Amazon and Facebook
greatly increased the number of CAHP and STP neighbors.

Similar analysis on transit providers revealed that they mostly interconnect with EC and CAHPs. These neighbors exhibit the largest reduction, indicating a market loss in their core business, probably impacting their dominance in the Internet connectivity.

\begin{figure}[tbh]
\centering
\epsfig{file=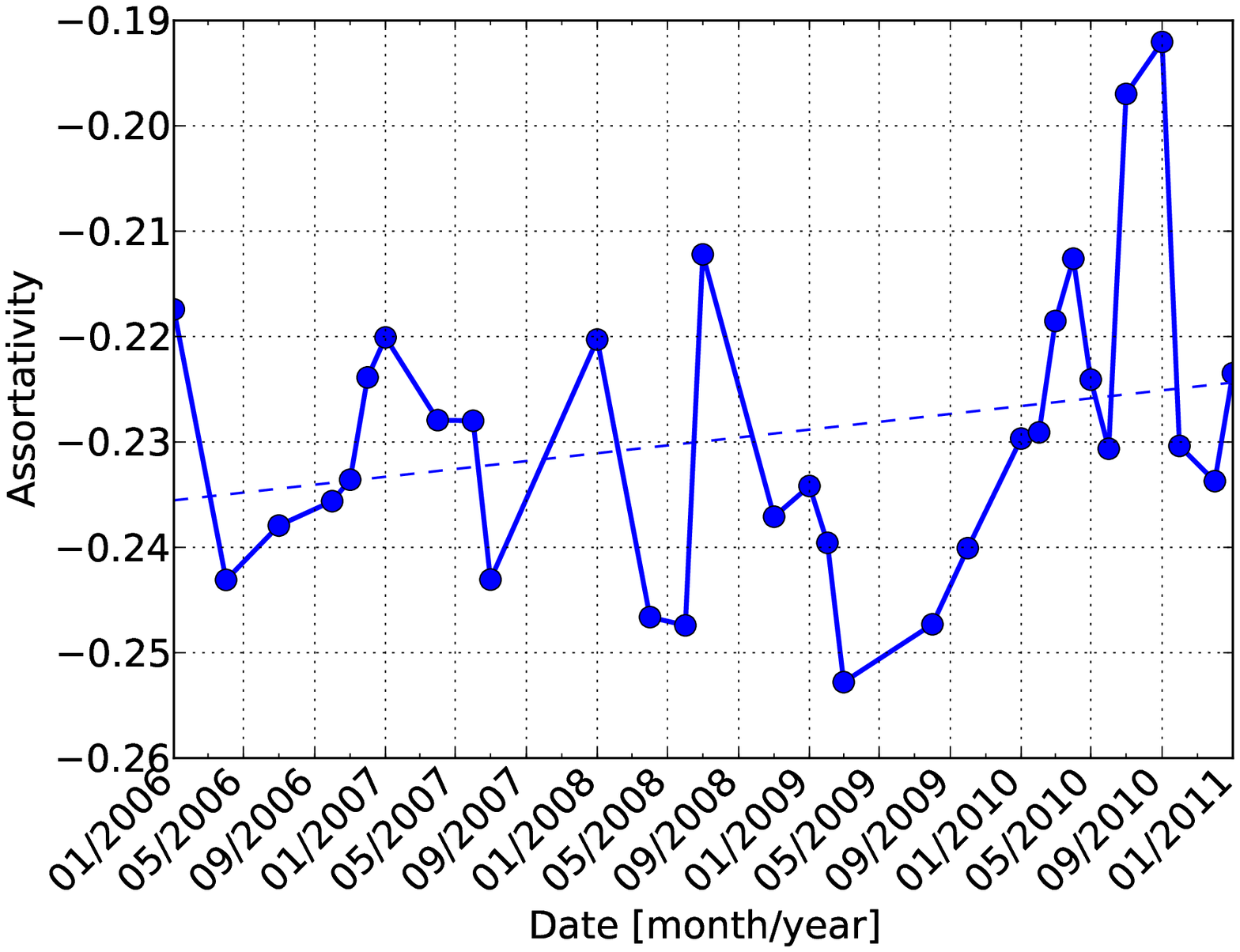,width=\figsize\textwidth}
\caption{Assortativity coefficient of the AS-graphs produced using DIMES dataset}
\label{fig:assortativity}
\end{figure}

Finally, we measure the assortativity coefficient values \cite{newman} of the complete AS graph, using data from DIMES. These
coefficients are positive when vertices tend to connect with similar-degree vertices and negative otherwise. The
AS-graph was shown to be disassortative \cite{newman,mahadevan}, meaning comprise mostly of radial links, connecting ASes towards the tier-1 ASes. Looking at the trend of the assortativity coefficient depicted in \figref{fig:assortativity} shows that the AS-graph becomes less disassortative, indicating that ASes increase
their connections with other similar-degree ASes.
% The graph here is very noisy - why not using iPlane data?

\ignore{
Finally, we measure the assortativity coefficient values \cite{newman} of the complete AS graph, using data from DIMES.
Positive coefficients mean that vertices tend to connect with similar-degree vertices and negative otherwise. The
AS-graph was shown to be disassortative \cite{newman,mahadevan}, meaning comprise mostly of radial links, connecting ASes towards the
tier-1 ASes. We found a slowly increasing trend (from -0.24 to -0.21), indicating that the AS-graph becomes less disassortative, i.e., more
ASes interconnect with other similar-degree ASes.
}

Overall, these findings agree with previous findings \cite{flatnet,labovitz,amogh2010}, and show a trend of large content providers that shift from
relying mostly on large transit
providers towards a flatter topology, interconnecting with smaller networks for gaining transit
and direct access to last-mile customers. Content providers that have been around longer, such as Google, Yahoo!, and MSN do not increase
the number of LTPs, mainly since they already exploit the benefit of connecting with them. The
``younger" content providers follow this trend and mostly use STPs and CAHP. This trend reveals an overall
decrease in the dominance
of tier-1 networks in the Internet ecosystem. This observation repeats in the following sections, when we
study the centrality of these networks.

%%[[TBD -- interesting to look at the 9 top tier-1 networks - how many of these does google connect? if it's only 1,2 then it's cool]]

\subsection{Geographical Connectivity}
We further measure how the spatial connectivity of the content providers evolves. Each neighboring AS is resolved to its major country, i.e.,
the country that hosts most of its known IP addresses,
using data from MaxMind \cite{maxmind}. This database is considered accurate in the country level of resolution \cite{maxmind-problem,noa_geolocation}.

\ignore{
%% Geographic distribution
\begin{figure}[tbh]
\centering
    \subfloat[Content]{
	\label{fig:country_content}
    \epsfig{file=fig/neighbor_country_content_real,width=\figsize\textwidth}
    }
    \hspace{-1mm}
    \subfloat[Transit]{
    \label{fig:country_transit}
    \epsfig{file=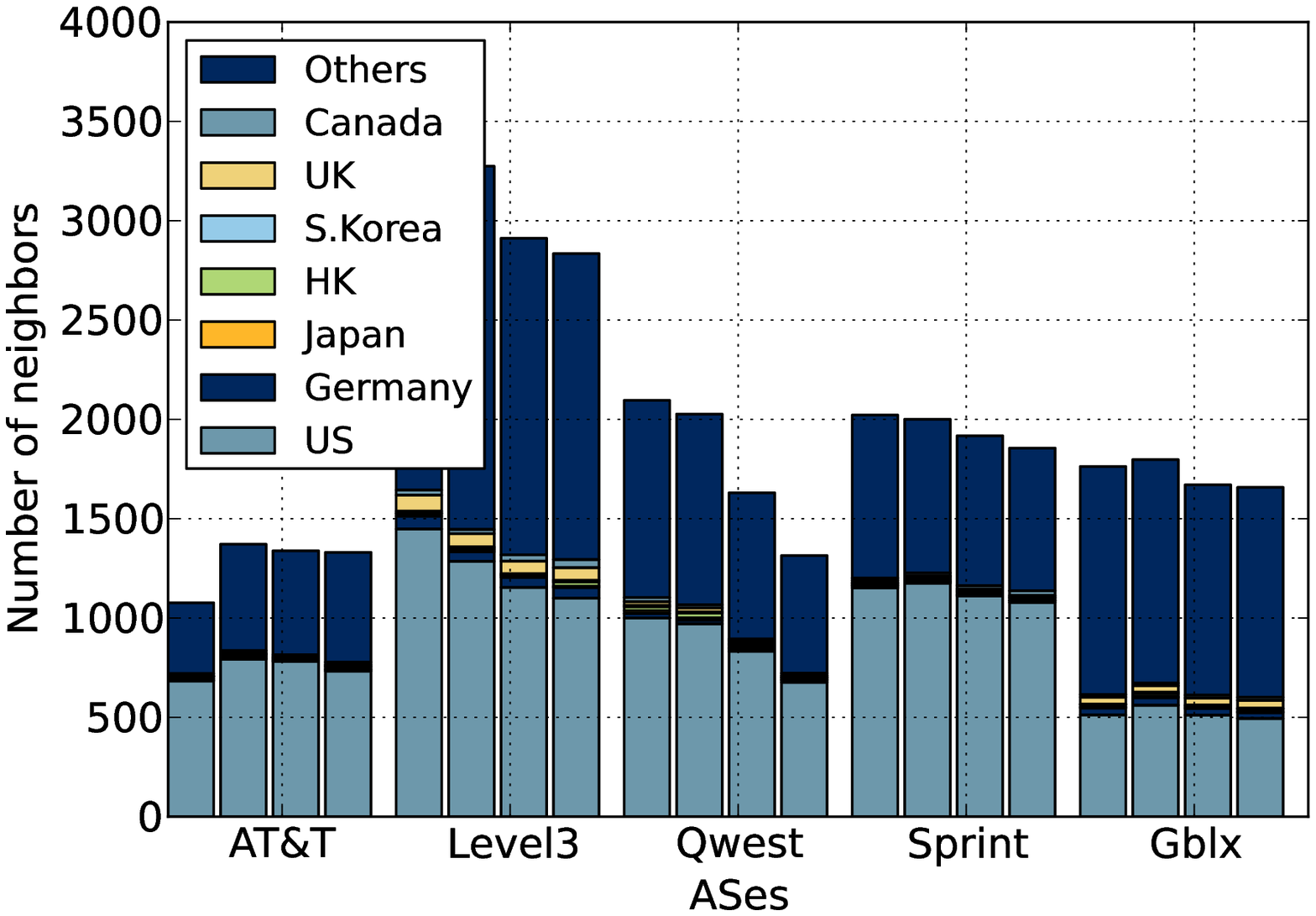,width=\figsize\textwidth}
    }
    \caption{Distribution of neighbor types}
    \label{fig:neighbor_country}
\end{figure}
}

\figref{fig:country_content} depicts, for each AS, the number of neighbors in a set of countries, over a period
of four years (the bars of each AS are for 2007, 2008, 2009, and 2010, from left to right).
The content providers we analyze originate from the US, hence it is not surprising that the largest percentage of
neighbors were found to be in the US. MSN, Google, and Yahoo! sustain a relative constant number of US-neighbors and
increase the number of non-US neighbors, whereas Amazon and Facebook, which grew faster, significantly increase the number of US-neighbors. AT\&T, Qwest and Sprint have a majority of neighbors in the US, whereas Level3 and Glbx have a lower percentage of US-neighbors. %The percentage of US-based neighbors decreases over time, since most new connections are made with non-US ASes.

\subsection{Density and Clustering}
The connectivity of the ASes in the graph directly affect its density and clustering.
Given a graph $G=(V,E)$, the density $D$ of the graph is defined as the number
of existing links out of the number of potential links, i.e.:
\begin{equation*}
D=\frac{2|E|}{|V|(|V|-1)}
\end{equation*}

\begin{figure}[tbh]
	\centering
	\epsfig{file=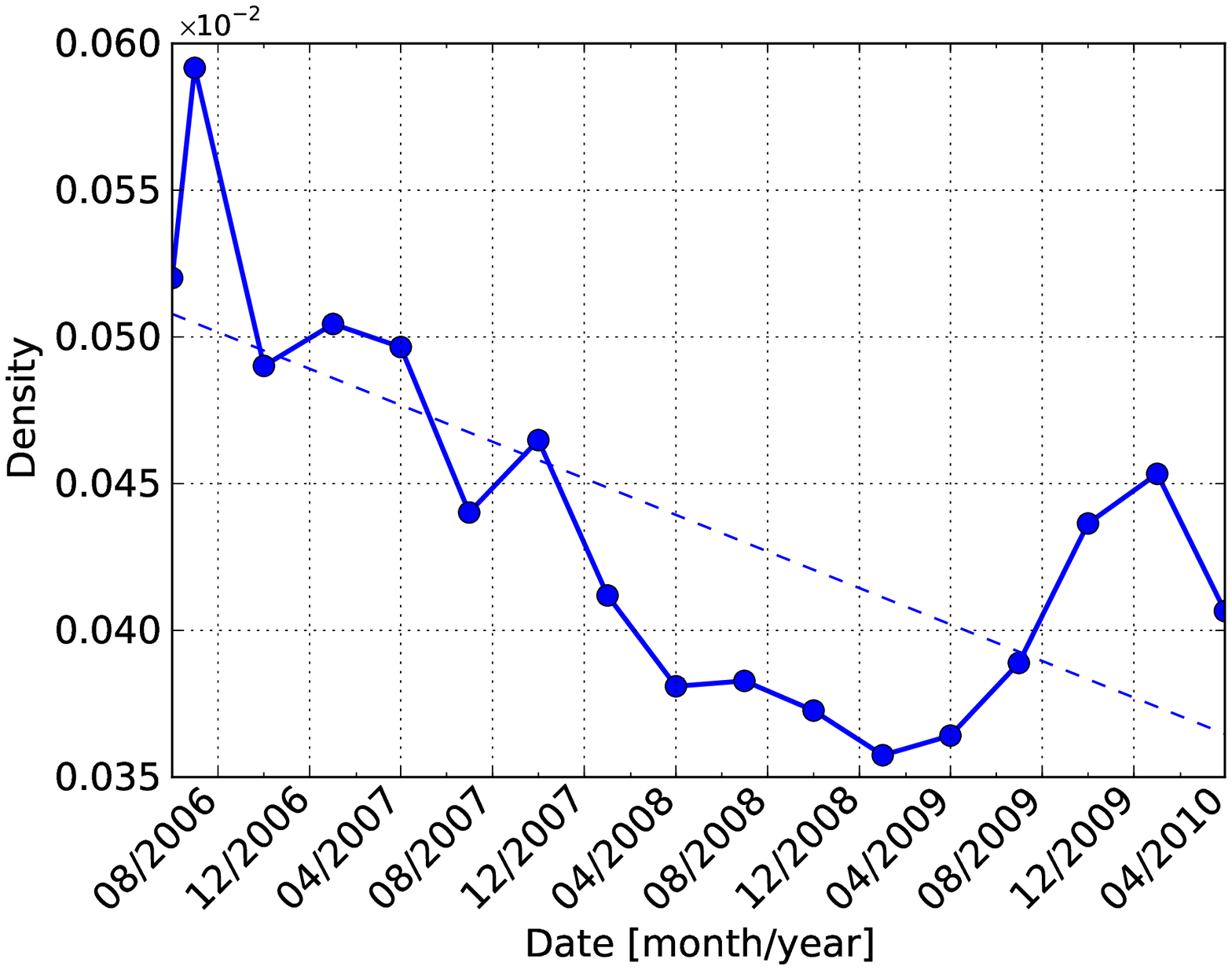,width=\figsize\textwidth}
    \caption{AS graph density (number of link out of the potential number of link)}\label{fig:density}
\end{figure}

\figref{fig:density} shows that the density of the AS-graph decreases over time, mainly since
new ASes that join the Internet significantly increase the potential of links, however, they
connect only to a small portion of the already existing ASes. This is expected since extremely large
degrees are observed only in a relatively small number of ASes, and even these connect to a few
thousands ASes.

In order to better understand the local connectivity of the ASes, we look at their
clustering coefficient (CC), which is a measure of the local density of an AS based on its neighbors. More formally,
the CC of a vertex in a graph is the number of triangles it forms with its immediate neighbors out of
the potential number of triangles.

\begin{figure}[tbh]
\centering
    \subfloat[Content providers]{
	\label{fig:cc_content}
    \epsfig{file=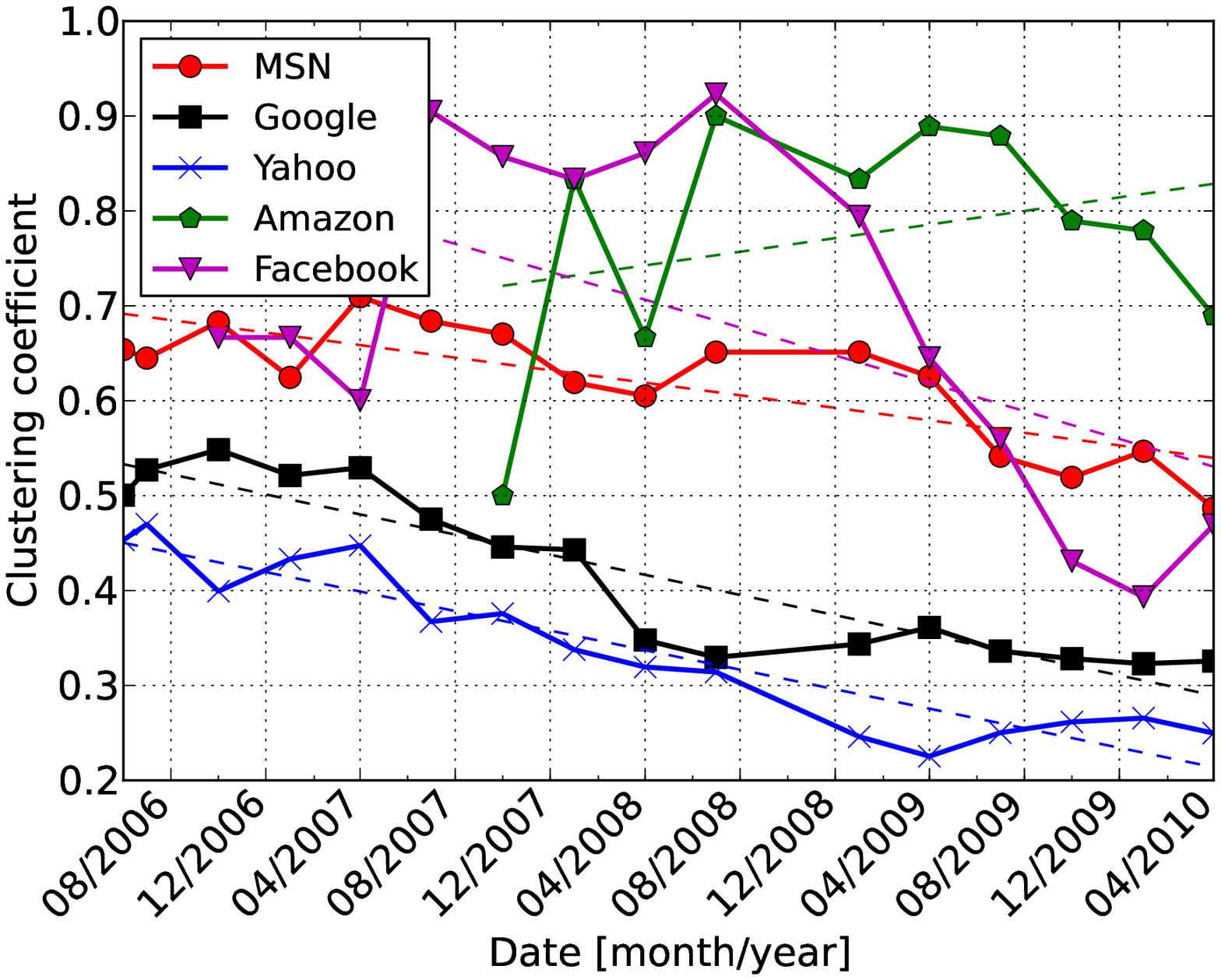,width=\figsize\textwidth}
    }
    \hspace{-1mm}
    \subfloat[Transit providers]{
    \label{fig:cc_transit}
    \epsfig{file=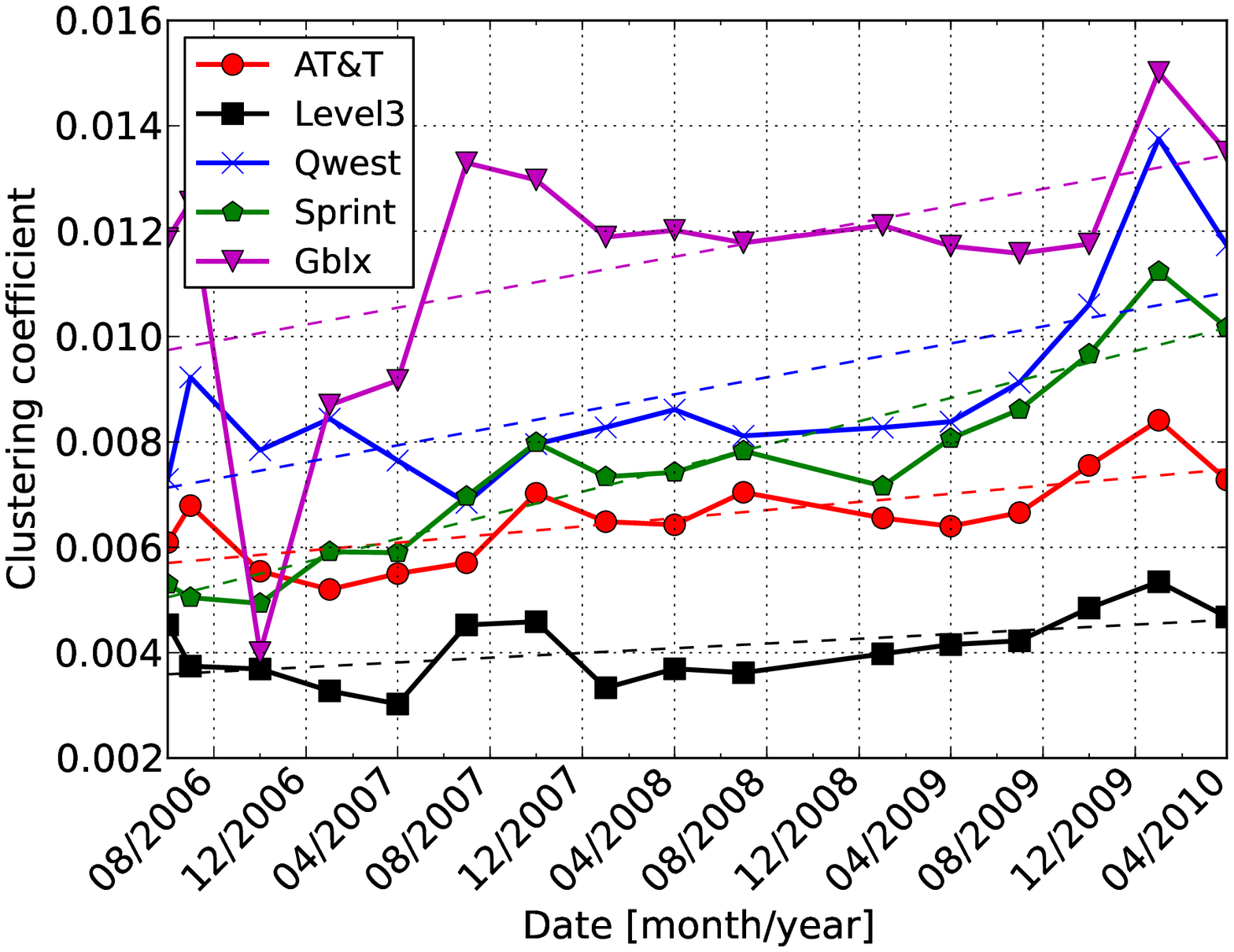,width=\figsize\textwidth}
    }
    \caption{Clustering coefficient}
    \label{fig:cc}
\end{figure}

\figref{fig:cc} depicts the CC of content and transit providers. \figref{fig:cc_content} shows that the CC of content providers is mostly decreasing, which
is a result of the increasing number of neighbors having few or no links amongst themselves. On the other hand, transit networks
exhibit an increasing CC, meaning that they are loosing neighbors with low connectivity and maybe hints at an increase 
in neighbor interconnections.

These two observations strongly
indicate that
content providers are increasing connectivity with access providers that are not in the core and are thus not interconnected. On the other hand,
transit networks lose small customers and the remaining customers increase their interconnections that are used to
bypass the core \cite{labovitz}, which causes the transit networks to gradually lose their centrality.

\subsection{IXPs}
IXPs are a convenient method for ASes to interconnect, since it provides a shared facility and infrastructure \cite{ixpsmapped}. ASes
that seek to expend their connectivity have an incentive to use such facilities as it enables
them to connect to a wide range of other ASes. As such, content providers can leverage IXP connectivity
to gradually increase the number of peering ASes with minimal setup costs.
In this analysis we use DIMES data since it manages to detect more IXP links than iPlane.

\begin{figure*}[tbh]
\centering
    \subfloat[Content providers]{
	\label{fig:ixps_content}
    \epsfig{file=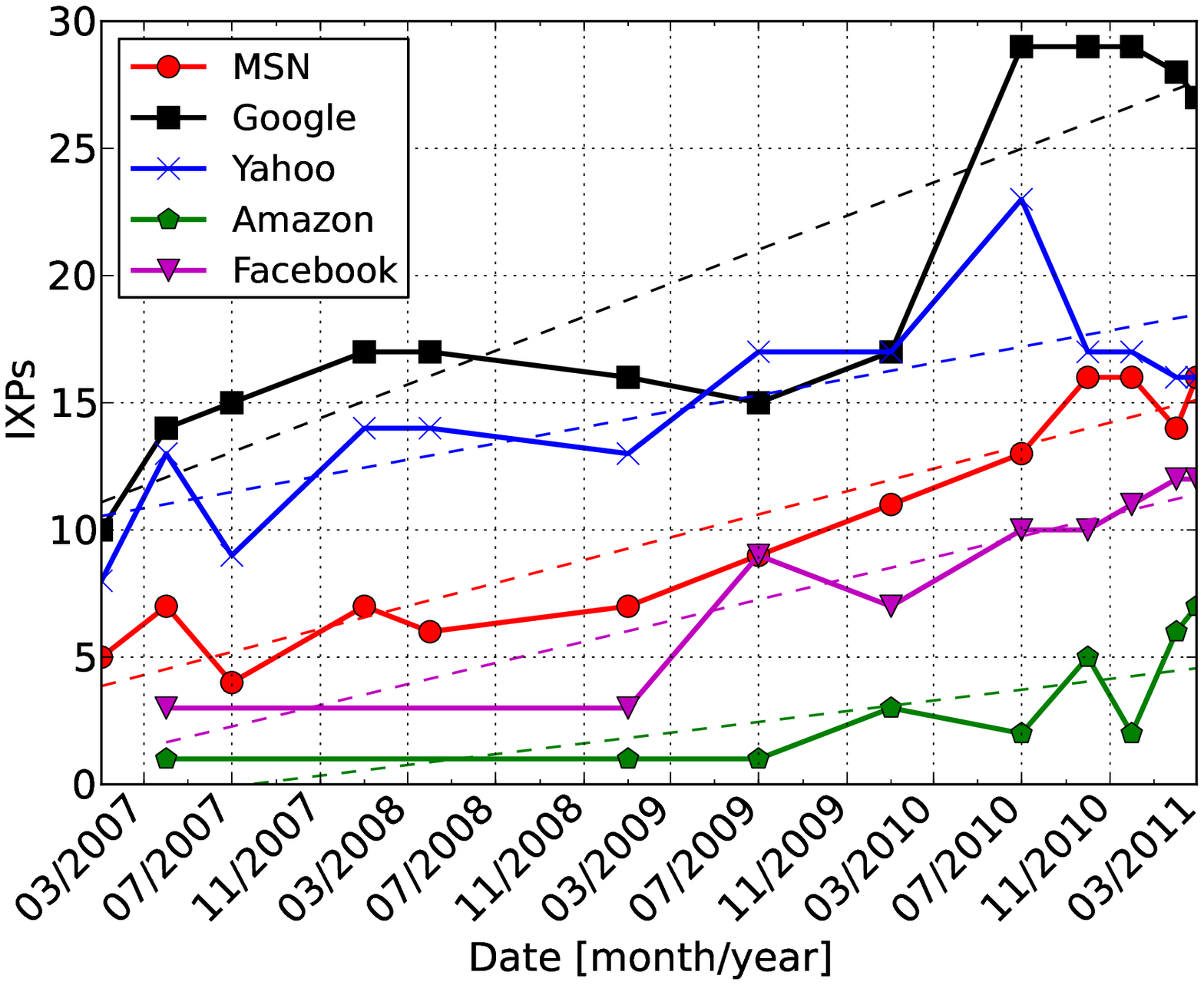,width=\figsize\textwidth}
    }
    \hspace{-1mm}
    \subfloat[Transit providers]{
    \label{fig:ixps_transit}
    \epsfig{file=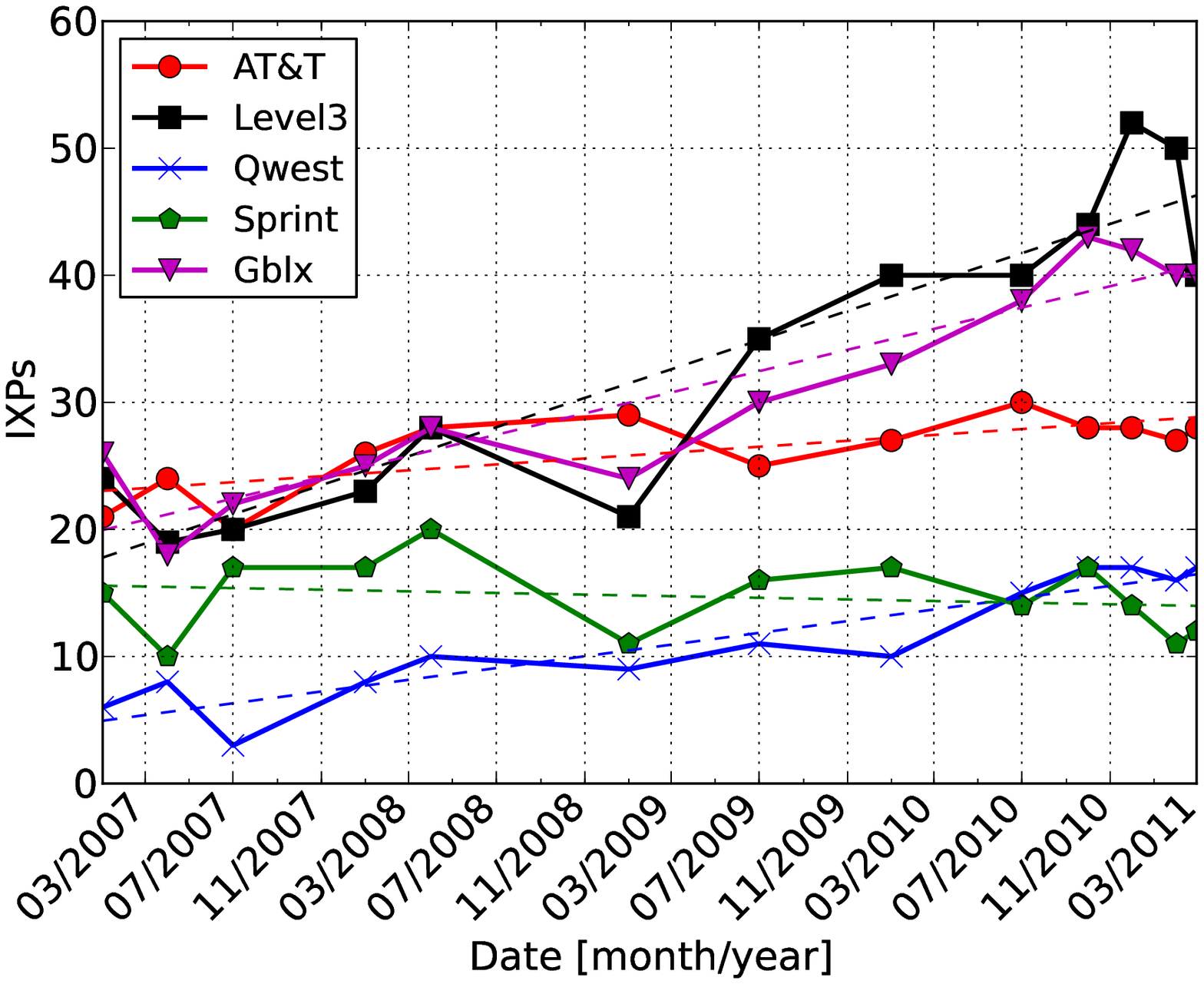,width=\figsize\textwidth}
    }
    \caption{Number of IXPs used by content and transit providers}
    \label{fig:ixps}
\end{figure*}

\figref{fig:ixps} shows the number of IXPs used by content and transit providers. \figref{fig:ixps_content} shows
that content providers gradually increased their adoption of IXPs mainly since early 2009. Most of
the content providers increased the number of IXPs they connect through by more than 100\% in only a few years, emphasizing
the important role that IXPs play in the Internet.

\figref{fig:ixps_transit} shows that transit providers use more IXPs, but their percentage out of the overall neighbors
is extremely low. Additionally, although transit providers increase their usage of IXPs, the number of IXPs they connect through grows much
slower than content providers.

\begin{figure*}[tbh]
\centering
    \hspace{-1mm}
    \subfloat[Content providers]{
	\label{fig:ixp_links_content}
    \epsfig{file=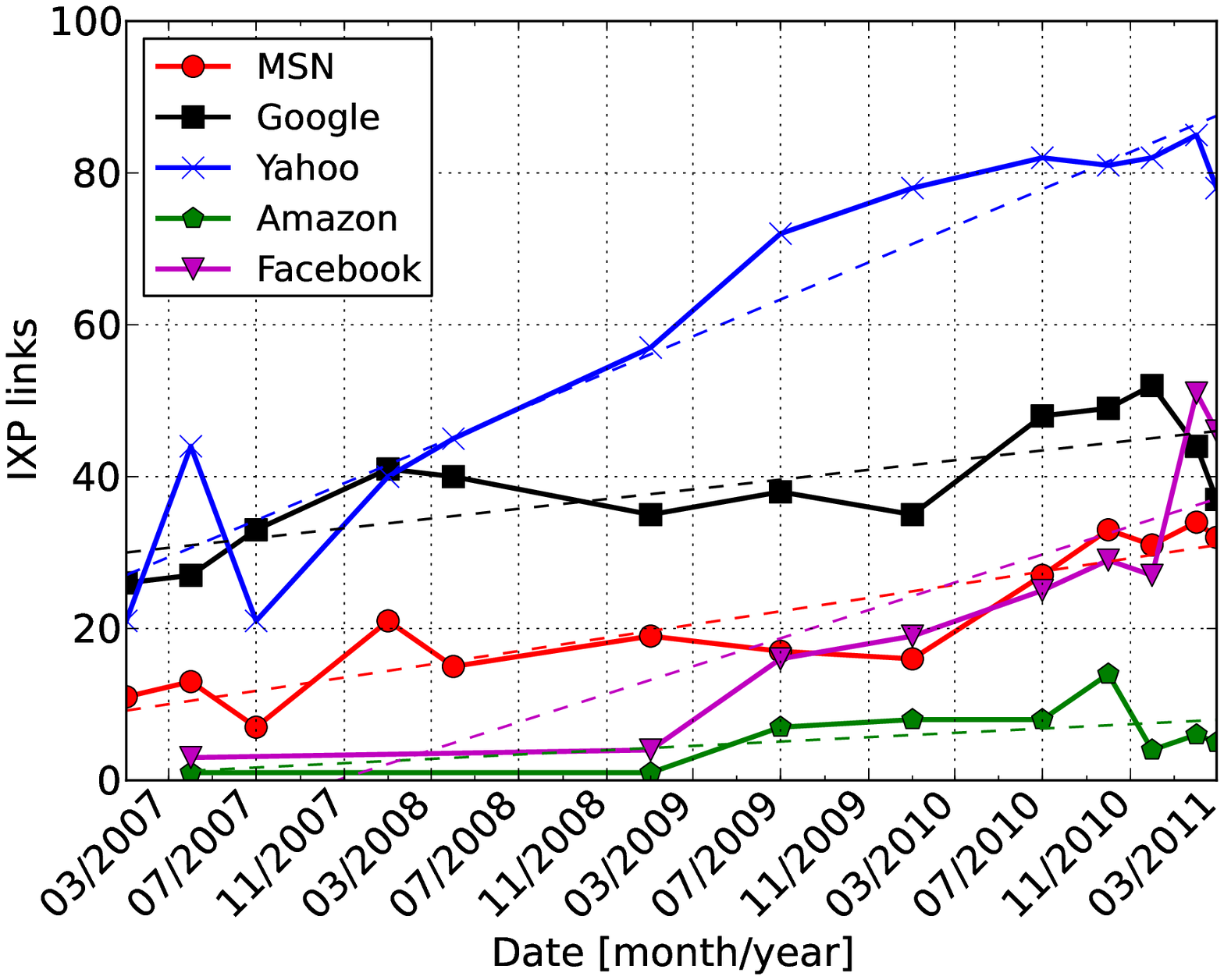,width=\figsize\textwidth}
    }
    \hspace{-1mm}
    \subfloat[Transit providers]{
    \label{fig:ixp_links_transit}
    \epsfig{file=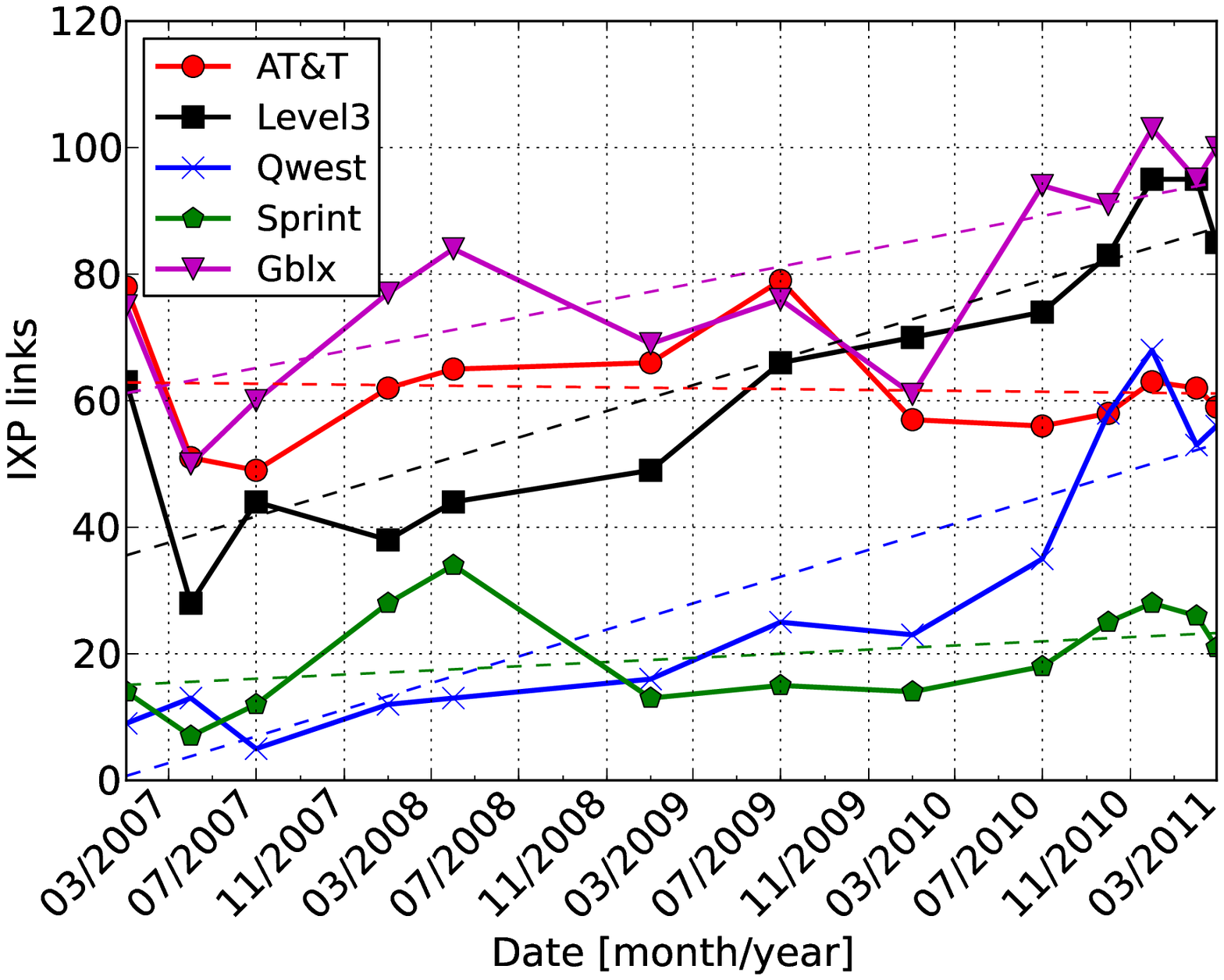,width=\figsize\textwidth}
    }
    \caption{Number of IXP links used by content and transit providers}
    \label{fig:ixp_links}
\end{figure*}

\figref{fig:ixp_links} depicts the number of AS-links that use an IXP, i.e., for each AS, the number of other ASes
that it connects through an IXP.
Although both transit and content providers exhibit a rising trend, the percentage of IXP links
used by content providers is significantly higher, reaching almost 40\% of their links, indicating
that content providers indeed embrace IXPs as method for increasing connectivity.

Interestingly, the number of IXP links is not significantly higher than the number of IXPs, meaning that
each IXP is used for connecting with only a few ASes. Since more ASes are expected to join these
shared facilities~\cite{ixpsmapped}, the growth potential of content networks connectivity is high.

\section{Autonomous Systems Hierarchy}
The hierarchical structure of the Internet has been studied extensively, classifying ASes into
the classical three-tiered model~\cite{ge01hierarchical,taxonomy}, and understanding the valley-free packet routing
rules~\cite{gao01}. A different method of hierarchical analysis is $k$-pruning~\cite{medusa-pnas},
which decomposes graphs into \emph{shells}, based on the node
connectivity towards the graph center. In the AS graph, ASes in the first shell are those who have only one link leading to the `center' of the graph, whereas
ASes in the $k$th shell have $k$-connectivity
towards the center. The \emph{nucleus} (or \emph{core}) is the shell with the highest index, which is
considered to contain top level providers, mostly tier-1 transit networks~\cite{medusa-pnas}.

\begin{figure}[tbh]
\centering
    \subfloat[Content providers]{
	\label{fig:shell_content}
    \epsfig{file=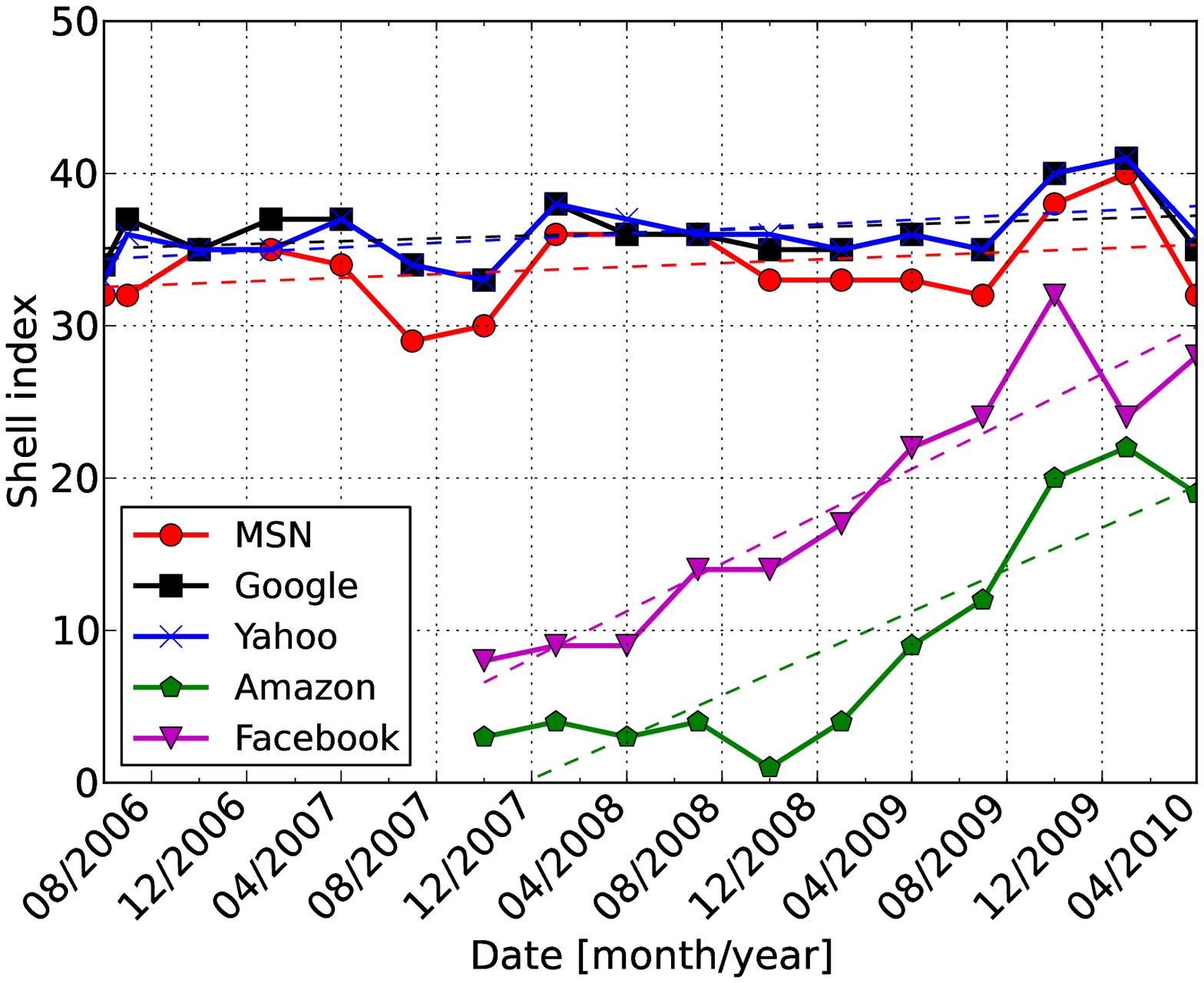,width=\figsize\textwidth}
    }
    \hspace{-1mm}
    \subfloat[Transift providers]{
    \label{fig:shell_transit}
    \epsfig{file=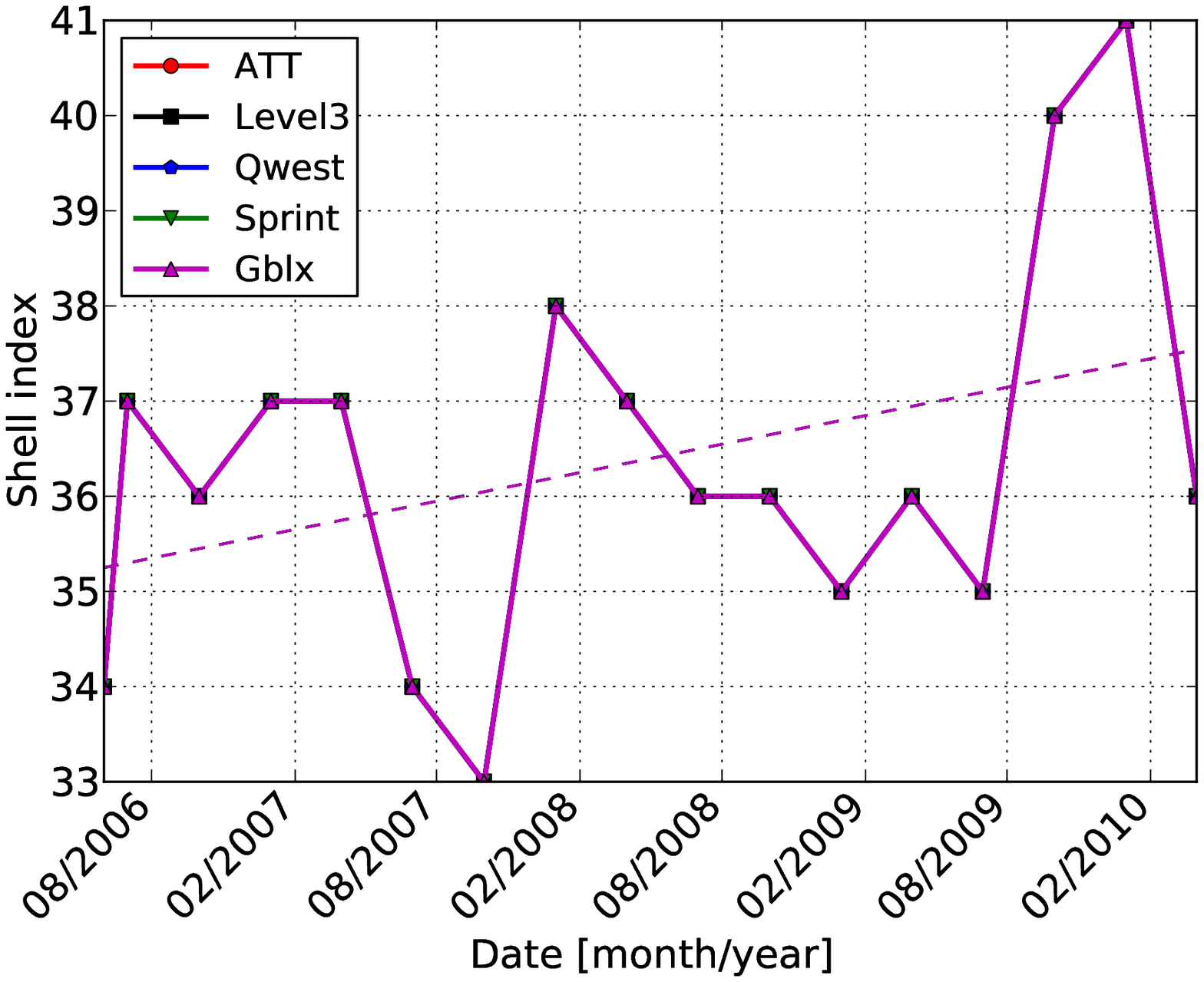,width=\figsize\textwidth}
    }
    \caption{Shell index of content and transit providers}
    \label{fig:shells}
\end{figure}

\figref{fig:shells} shows the shell index of content and transit providers.
\figref{fig:shell_transit} shows that
all of the examined transit networks are in the same shell, which is the nucleus. This
is expected, since these ASes are the top-level providers of the Internet, making
them extremely central~\cite{medusa-pnas,quantify-jsac}.

\figref{fig:shell_content} shows that Google,
Yahoo!, and MSN have a very high shell index, and are either in the nucleus or in a very close shell.
Facebook and Amazon exhibit a dramatic increase
in their shell index. %, reaching almost to the nucleus.
Facebook enjoyed increasing popularity, which drive it into connecting with high tier
networks, improving its connectivity and service levels. Amazon's increase reveals the
market shift that the company had, from being an online store that resides in a low shell into
a major cloud service provider, hosting numerous application. This change mandates significantly more
connections with tier-1 networks, resulting in the increase witnessed in 2009.  % change 2008 to 2009 - Yuval

For a content provider, being close to the nucleus has direct capital effect, because the higher
the shell a network is located in, the less it needs to be a paying customer of transit networks \cite{gao01}, mainly
since it does not need transit services in order to reach most other customer networks.
Looking back at the degree of content providers in \figref{fig:degree}, their degree is significantly lower than tier-1 transit networks.
Since tier-1 networks are expected to comprise the nucleus \cite{medusa-pnas}, it is unexpected that content providers
manage to reach high shells. However, examining the types of ASes in the nucleus, reveals that unlike previously thought \cite{medusa-pnas}, there
is a significant portion of STPs (roughly 40\%) and even CAHPs (15-25\%) in the nucleus. As content providers
increase the portion of neighbors that reside in the core, they manage to increase their shell index, thus
reduce their dependency on top-tier transit providers.
We note that the nucleus index has slightly increased since 2006 (roughly 5\% increase in iPlane and
10\% in DIMES). An increase in this index indicates a richer interconnection within the Internet core.

\section{Measuring Centrality}
In this section we seek to understand the ways that the changing connectivity of content providers
affects their centrality, and whether these changes actually change the centrality of large transit providers.
We approximate the potential load on the ASes, and use a web-based
centrality metric for assessing the centrality of the networks.

\subsection{Approximation of Load}
Approximating the potential load on an AS is an estimate of its importance, since it indicates
that an AS serves many other ASes for routing packets \cite{mahadevan}. This approximation is commonly achieved using betweenness centrality (BC).
In graph theory, BC measures the number of appearances of a vertex in the shortest paths between all other vertices,
relative to the total number of shortest paths. Given a vertex $v \in V$, its bc $B(v)$ is calculated using:

$$B(v) = \sum_{s \neq v \neq t \in V} \frac{\sigma_{st}(v)}{\sigma_{st}}$$

\noindent where $\sigma_{st}$ is the number of shortest paths from $s$ to $t$ and
$\sigma_{st}(v)$ is the number of shortest paths from $s$ to $t$ that pass through $v$.

Usually, BC is normalized by dividing it by the number of possible pairs,
to enable comparison between graphs of different sizes. Given
the number of vertices in the graph $n=|V|$, in undirected graphs the number of pairs
is $n(n-1)$.

% I changed the text below.  Instead saying these are two methods that can be used I wrote it as: the common way is wrong, lets quantify the error, and use it because the error is small.  It puts us in a better light.

BC is commonly applied to the AS-graph for measuring
the possible load that an AS sustains. However, since packets traverse the Internet in valley-free paths \cite{gao01},
BC can only serve as an approximation for node centrality. To check the validity of this approximation, we 
first calculated the BC on the undirected AS graph, ignoring valley-free paths.
Then, in order to account to the valley-free rules, we calculated the BC of each AS directly from the probed paths of each month, by dividing
the number of traceroutes that traverse an AS (in the middle of the trace) by the total number of traceroutes. This method
is more accurate than adding directions to the AS-graph, since inferring commercial relationships between ASes was
shown to contain errors, with up to 20\% mistakes in peer-to-peer relationships \cite{XiaGao04,dimitropoulos-2006-37}. 
We found that the measurement inference of centrality gave similar BC values
as calculating shortest path on the undirected AS graph, with the measurement inference exhibiting slightly lower BC values %(because not all shortest paths are valley-free, thus do not appear) 
and significantly more noise. Since we are interested in trends where the exact
values are less significant, we
use the common and easier method, which is less accurate but provides a clear view of the trends.

\begin{figure}[tbh]
\centering
    \subfloat[Content providers]{
	\label{fig:bc_content}
    \epsfig{file=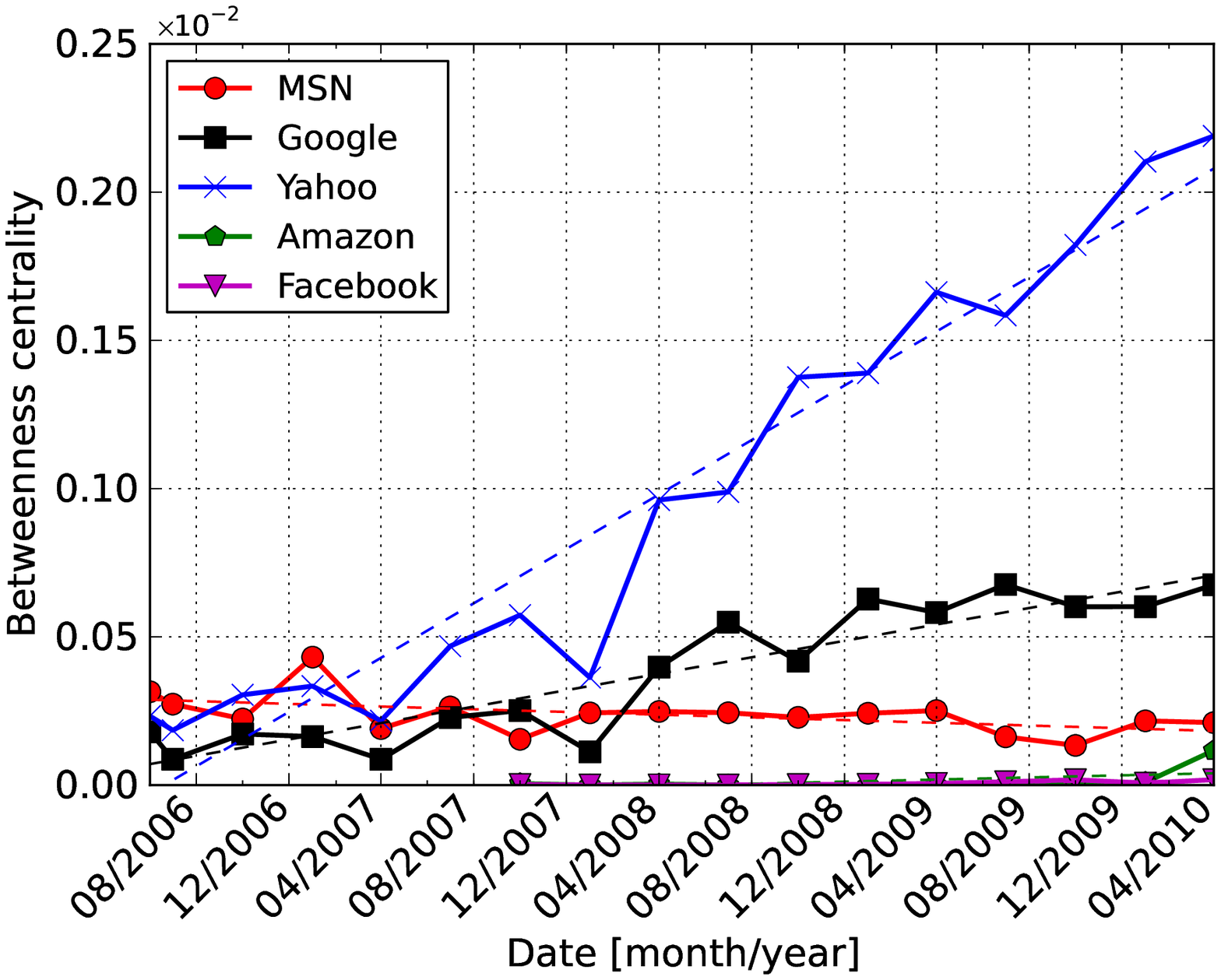,width=\figsize\textwidth}
    }
    \hspace{-1mm}
    \subfloat[Transit providers]{
    \label{fig:bc_transit}
    \epsfig{file=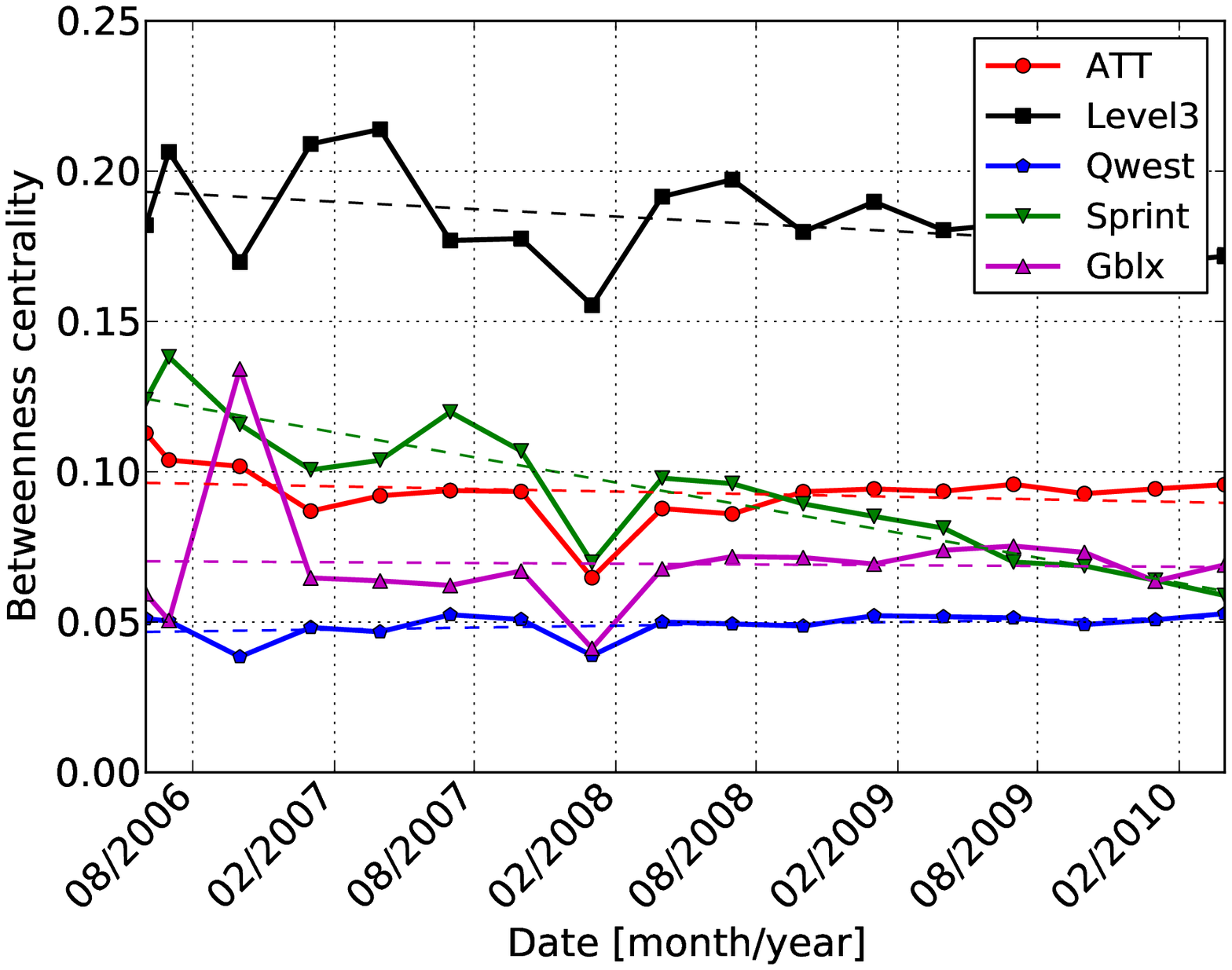,width=\figsize\textwidth}
    }
    \caption{Betweenness centrality}
    \label{fig:bc}
\end{figure}

\figref{fig:bc} shows the normalized BC of the selected content and transit networks. As expected,
the BC values of the tier-1 transit networks are significantly higher (two orders of magnitude)
than content network, as the latter are usually the last hop of the routes.

Interestingly, \figref{fig:bc_content} shows an increasing trend in the BC of Google and Yahoo!.
We validated this increase with the DIMES dataset and observed the same trend, with the
difference that Google witnesses a slightly higher increase. We followed the
AS-level traces of Yahoo! and Google during weeks 39 and 40 of 2010, and found that
Yahoo! AS10310 and Google AS15169 appear mid-trace in roughly 22\% and 14\% of
the traces they appear in, respectively. We further looked at the traces themselves,
and found that both networks always appear as the hop-before-last, and are siblings
of the last hop (owned by either Google or Yahoo!). In Google's case, the majority of traces terminate with YouTube (AS36561).
Yahoo! on the other hand is comprised of several different ASes,
such as Yahoo! Japan (AS23926), Yahoo! US (AS7233), and Yahoo! Backbone (AS24018).

These two networks exhibit a similar behavior: they both provide transit for sibling ASes as well as content, but
only for data belonging to their networks.
Yahoo!'s major
AS does this because their network is comprised of several regional and probably specialized networks.
Google provides transit to companies they purchase, leveraging their major AS's connectivity.
In both cases, the end result is the same -- content providers, which were once mainly stub networks that
terminate routes, are becoming more central by providing transit towards other sibling and peering ASes,
enabling network operators to save costs by leveraging their gradually improved connectivity. This
observation is also backed from the findings of Labovitz~\etal~\cite{labovitz}, who showed a decreasing
traffic trend observed from YouTube AS and an increasing traffic trend from Google's AS. This is the result
of Google's AS acting as the major transit provider for YouTube's traffic.

%% TBD -- we should also look at the actual paths, since shortest paths are only an approximation. Analyze the
%% paths in the past and today, and show some bc values (no need for a plot).

\subsection{PageRank Centrality}
PageRank (PR), which was proposed initially for scoring web-pages~\cite{pagerank},
is a measure of the importance of a vertex in a network~\cite{Mehta06rankingattack,Liu07videosearch}. PR assigns relative
scores to all vertices in the network based on the iterative principle that high-scoring neighbors
contribute more to the score of a
vertex than low-scoring neighbors. More formally, given a graph $G=(V,E)$, the PR of a vertex $u \in V$ is iteratively computed
using:
\begin{equation*}
PR(u)=\displaystyle \sum_{v\in V,~(v,u) \in E} \frac{PR(v)}{degree(v)}
\end{equation*}

\noindent where the initial $PR$ of all vertices is set uniformly, meaning: $\forall u \in V,~PR(u)=1/|V|$. Notice
that this is a very simple formulation of PR, and a more complete form was used, however, the exact details are
not essential for understanding its meaning.

\begin{figure}[tbh]
\centering
    \subfloat[Content providers]{
	\label{fig:pr_content}
    \epsfig{file=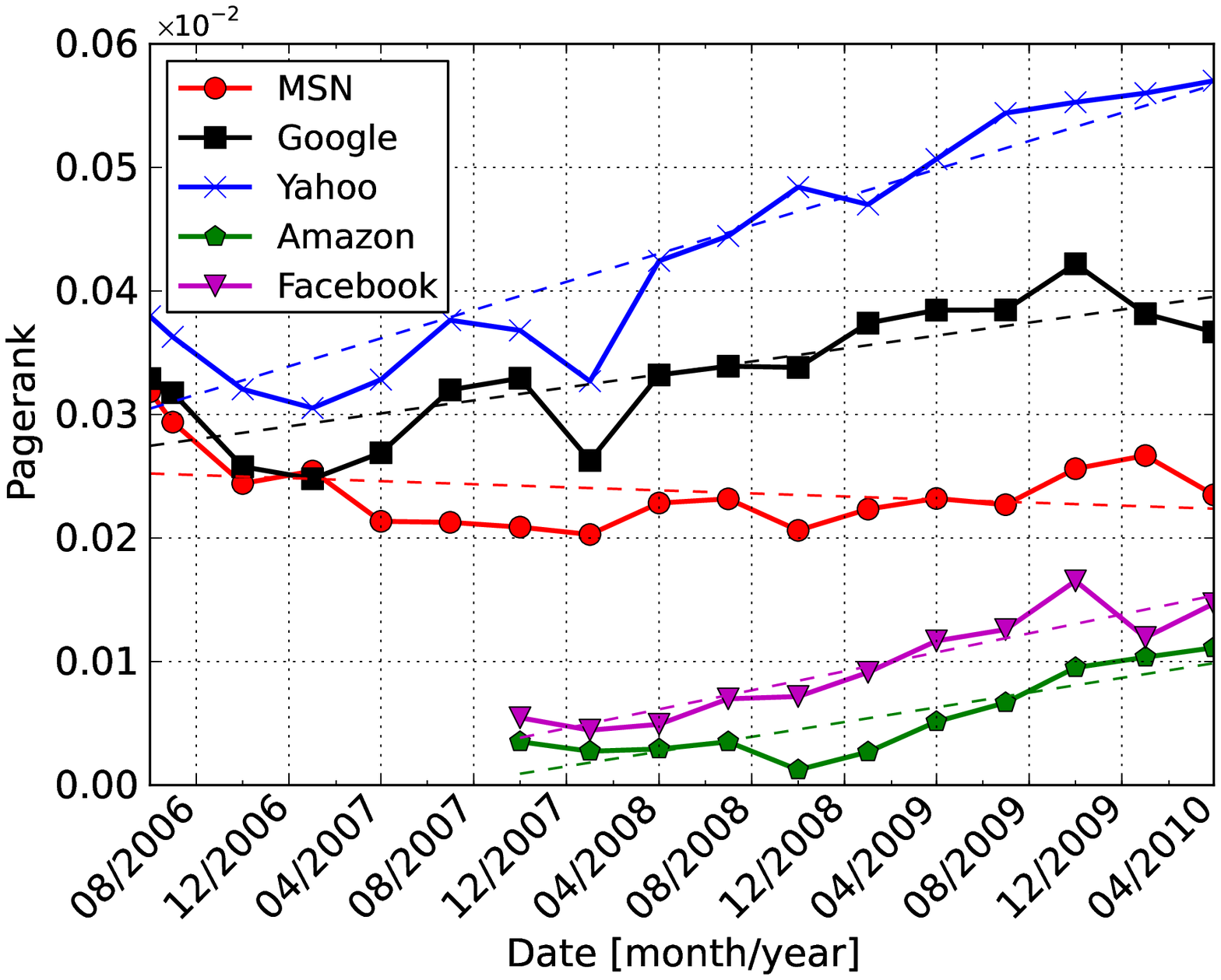,width=\figsize\textwidth}
    }
    \hspace{-1mm}
    \subfloat[Transit providers]{
    \label{fig:pr_transit}
    \epsfig{file=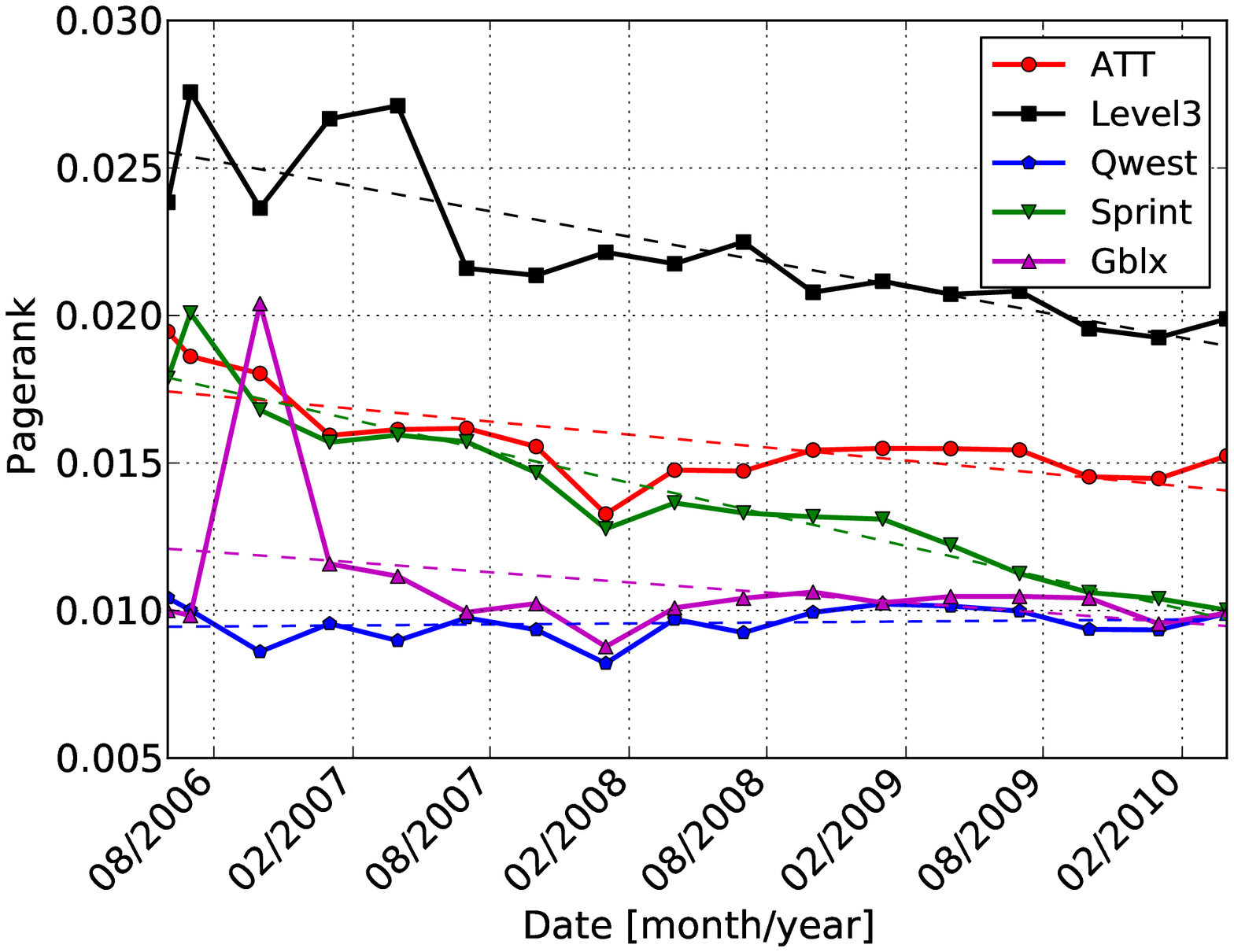,width=\figsize\textwidth}
    }
    \caption{PageRank centrality}
    \label{fig:pr}
\end{figure}

We applied PR on the AS graphs for extracting the centrality of ASes. \figref{fig:pr} shows the PR score of content and transit providers, strengthening the above observations.
The PR of transit providers is two orders of magnitude larger than content providers, indicating
that they are significantly more central than content providers. However,
while the PR of transit networks slowly decreases, content providers witness a gradual increase, with
MSN being an exception.

\section{Discussion} The AS-level graph exposes the trend that is changing the Internet -- content providers
become key players in the Internet. The connectivity trends indicate that content providers increase
the number of neighbors and diversify their types and geographical spread. They mostly
make new connections with small transit and access providers, enabling them a better reach
to worldwide customers, while minimizing high-tier transit costs. These changes
are also witnessed in the way content providers are gradually climbing towards the core of the Internet,
actually reforming it so that it includes non-transit networks.

IXPs are also gaining popularity, and they seem to be underutilized, meaning that the growth potential
using peering agreements is high. As they provide an easy method for interconnecting with other networks,
they are likely to be better utilized in the future.
%% Interesting to see the trend in the number of ASes per IXP

The centrality of content providers is a strong indication to their increasing dominance, and on the other hand,
the changing role of top-tier transit providers. The latter, despite offering attractive wholesale pricing plans \cite{tiers},
lose their dominance, as content providers, which are amongst the highest paying customers due to the
amount of traffic they produce~\cite{labovitz}, find ways to reduce redundancy on tier-1 networks, while better
utilizing their own resources. In addition to saving transit costs, content providers increase their offerings
to customers by providing Software as a Service (SaaS), such as hosting, cloud services, collaboration tools and others,
without worrying for network neutrality violations or other traffic shaping that can interfere with the service grade they provide~\cite{amogh2008,beverly2007,udi2011}.

\section{Conclusion}  This paper studies the topological trends of large content providers in the AS-level graph. We show
that these networks increase their connectivity with other networks, mostly using IXPs. As a result, the centrality of these
networks increase, as they manage to reduce their dependency in tier-1 transit networks, which lose dominance in
the Internet ecosystem. Through well-designed selection of peers, usage of IXPs, and acting as transit networks for traffic
originating from sibling ASes, the large content providers manage to
climb the Internet hierarchy, enabling them to improve service to their customers while cutting down transit costs.

\bibliographystyle{abbrv}
\bibliography{providers}

% that's all
\end{document}